\documentclass[useAMS,usenatbib]{mn2e}
\usepackage{graphicx,color,oldgerm}
\usepackage{amssymb,amsmath}
\usepackage[breaklinks=true]{hyperref}

\let\sun=\odot

\newcommand{\MBH}{\ensuremath{M_{\mathrm{BH}}}}
\newcommand{\Msun}{\ensuremath{{\rm M}_{\odot}}}
\newcommand{\Mstar}{\ensuremath{M_{\ast}}}
\newcommand{\Nstar}{\ensuremath{N_{\ast}}}
\newcommand{\Rstar}{\ensuremath{R_{\ast}}}
\newcommand{\Rnb}{\ensuremath{R_{\rm NB}}}
\newcommand{\Npart}{\ensuremath{N_{\rm p}}}

\newcommand{\Rsun}{\ensuremath{{\rm R}_{\odot}}}

\newcommand{\pccube}{\ensuremath{\mathrm{pc}^{-3}}}

\newcommand{\kms}{\ensuremath{\mathrm{km\,s}^{-1}}}

\newcommand{\Vrel}{\ensuremath{V_{\rm rel}^{\infty}}}
\newcommand{\Vstar}{\ensuremath{V_\ast}}
\newcommand{\trh}{\ensuremath{t_{\rm rh}}}
\newcommand{\trc}{\ensuremath{t_{\rm rc}}}
\newcommand{\tcc}{\ensuremath{t_{\rm cc}}}

\newcommand{\tstar}{\ensuremath{t_\ast}}
\newcommand{\tcoll}{\ensuremath{t_{\rm coll}}}

\newcommand{\trlx}{\ensuremath{t_{\rm rlx}}}

\newcommand{\Rcore}{\ensuremath{R_{\rm core}}}
\newcommand{\Mcore}{\ensuremath{M_{\rm core}}}

\newcommand{\Rh}{\ensuremath{R_{\rm h}}}
\newcommand{\GCoulomb}{\ensuremath{\gamma_{\rm c}}}
\newcommand{\sigmaOneD}{\ensuremath{\sigma_{v}}}
\newcommand{\sigmaThreeD}{\ensuremath{\sigma_{3}}}
\newcommand{\MESSY}{ME(SSY)**2}
\newcommand{\SPEDI}{SPEDI}
\newcommand{\NBFOUR}{NBODY4}

\newcommand{\rem}[1]{} 
\newcommand{\Comment}[3]{}

\newcommand{\Sim}[1]{\texttt{#1}}

\title[Runaway collisions]{%
Runaway collisions in young star clusters. I.~Methods and tests}

\author[M. Freitag, F. A. Rasio and H. Baumgardt ]{
Marc Freitag$^{1,2}$\thanks{Present address: Institute of Astronomy, University of Cambridge, Madingley Road, Cambridge CB3~0HA, UK. E-mail: freitag@ast.cam.ac.uk},
Frederic A.\ Rasio$^2$
and Holger Baumgardt$^3$ 
\\
$^1$Astronomisches Rechen-Institut, M\"onchhofstrasse 12-14,
D-69120 Heidelberg, Germany\\
$^2$Department of Physics and Astronomy, 
Northwestern University, Evanston, IL 60208, USA\\
$^3$Sternwarte, Universit\"at Bonn, Auf dem H\"ugel 71, 53121 Bonn, Germany
}

\begin{document}

\date{Accepted. Received; in original form}

\pagerange{\pageref{firstpage}--\pageref{lastpage}} \pubyear{2005}

\maketitle

\label{firstpage}

\begin{abstract}
We present the methods and preparatory work for our study of the
collisional runaway scenario to form a very massive star (VMS,
$\Mstar>400\,\Msun$) at the centre of a young, compact stellar
cluster. In the first phase of the process, a very dense central core
of massive stars ($\Mstar\simeq30-120\,\Msun$) forms through mass
segregation and gravothermal collapse. This leads to a collisional
stage, likely to result in the formation of a VMS (itself a possible
progenitor for an intermediate-mass black hole) through a runaway
sequence of mergers between the massive stars. In this paper we
present the runaway scenario in a general astrophysical context. We
then explain the numerical method used to investigate it. Our approach
is based on a Monte Carlo code to simulate the stellar dynamics of
spherical star clusters using a very large number of particles (a few
$10^5$ to several $10^6$). Finally, we report on test computations
carried out to ensure that our implementation of the important physics
is sound. In a second paper, we present results from more than 100
cluster simulations realized to determine the conditions leading
to the collisional formation of a VMS and the characteristics of the
runaway sequences.
\end{abstract}


\section{INTRODUCTION}

Runaway collisions and mergers of massive stars following gravothermal
contraction and core collapse in a young, dense star cluster provides
a natural path to the formation of a massive object at the centre of
the system (\citealt{EbisuzakiEtAl01}; \citealt{PZMcM02};
\citealt*{GFR04}, hereafter GFR04). The basic idea goes back to the
earliest studies of the quasar/AGN phenomenon
\citep[e.g.,][]{SS66,Colgate67,Sanders70b}.

Runaways could easily occur in a variety of observed young star
clusters such as the ``young populous clusters'' (like the Arches and
Quintuplet clusters near our Galactic centre) and the ``super star
clusters'' found in all starburst environments, including most
galactic mergers \citep{FMcLM99,GS99}. The Pistol Star observed in the
Quintuplet cluster \citep{FNMMcLGGL98} may well be the product
of such a runaway, as demonstrated by direct $N$-body simulations
\citep{PZMcM02}. If the massive runaway collision product collapses
to a black hole (BH), this scenario also provides a route to the
formation of intermediate-mass black holes (IMBH) in star
clusters. Dynamical evidence for IMBHs at the centres of some globular
clusters has been reported for many years
\citep{GRH02,GerssenEtAl02,vdMEtAl02}.  Tentative evidence has
also been reported for an IMBH in the Galactic centre source IRS~13,
which was recently resolved into a small cluster of bright stars
\citep{MPSR04}. This could be the remnant of a much larger cluster
that got tidally disrupted as its orbit around the Galactic centre
decayed through dynamical friction \citep{HM03,GR05}. The very bright
ultra-luminous X-ray source (ULX) associated with the young star cluster
MGG~11 in the starburst galaxy M82 could also have been produced
through runaway collisions
\citep{PZBHMM04}. A similar process may be responsible for the
formation of IMBHs in larger clusters, such as proto-globular
clusters, as well as seed BHs in proto-galactic nuclei. These can
later grow through a variety of processes including gas accretion,
stellar captures, and mergers \citep{MCD91,FB02b,YT02,HM03,WL03,Blandford04}.

The questions we address here are also central to our
understanding of low-frequency gravitational-wave (GW) sources and the
development of data analysis and detection strategies for these
sources. In addition, our work may be viewed as
mainly relevant to galactic astronomy, the results could also have
applications in extragalactic astronomy and cosmology.  The direct
injection into the centre of a galaxy of many IMBHs produced by
collisional runaways in nearby young star clusters provides an
important new channel for building up the mass of a central
supermassive BH through mergers \citep{PZMcM02}.  It is possible that
this process is still ongoing in our own Galactic centre
\citep{HM03,KFM04,GR05}. In contrast, minor
mergers of galaxies are unlikely to produce BH mergers, as the smaller
BH will rarely experience enough dynamical friction to spiral in all
the way to the centre of the more massive galaxy \citep{VHM03}. These
ideas are also of critical importance for the design and planning of
LISA, since the inspiral of an IMBH into a SMBH provides one of the best
sources of low-frequency GWs for direct study of strong field gravity
with a space-based interferometer \citep{CT02,Phinney03,CH04,Miller05}.
Although the SMBHs found in bright quasars and many nearby galactic
nuclei are thought to have grown mainly by gas accretion 
\citep[e.g.,][]{Soltan82,HNR98,FI99,Richstone04},
current models suggest that LISA will probe most efficiently a
cosmological massive BH population of lower mass, which is largely
undetected \citep{Menou03}. LISA will measure their masses with
exquisite accuracy, and their mass spectrum will constrain formation
scenarios for high-redshift, low-mass galaxies and, more generally,
hierarchical models of galaxy formation
\citep[e.g.,][]{HK02,HughesHolz03,VHM03,SHMV04}.

GFR04 concentrated on the early dynamical evolution of young, dense
star clusters. They performed dynamical Monte Carlo simulations for systems
containing up to $10^7$ stars, and followed the rapid mass segregation of
massive main-sequence (MS) stars and the development of the Spitzer
instability. They showed that, with a realistic initial mass function (IMF),
these systems can evolve to core collapse in a small fraction of the
initial half-mass relaxation time. If the core-collapse time is less
than the lifetime of the most massive MS stars, all stars in the collapsing
core may then undergo runaway collisions. The study in GFR04 was limited
to the first step in this process, up to the occurrence of core collapse.
About 100 simulations were performed for clusters with a wide variety of
initial conditions, varying systematically the cluster density profile,
stellar IMF, and the number of stars. GFR04's results confirmed that, for
clusters with a moderate initial central concentration and any realistic
IMF, the ratio of core-collapse time to initial half-mass relaxation time
is typically $\sim 0.1$, in agreement with previous calculations. It was
also found that, for all realistic initial conditions, the mass of
the collapsing core (at the onset of collapse) is always close to
$\sim10^{-3}$ of the total cluster mass, very similar to the observed
correlation between central BH mass and total cluster mass in a variety
of environments \citep{Magorrian98,MF01,HR04}.

In this and a following paper (\citealt*{FGR05}; hereafter Paper~II),
we go a step further in our study of runaways, by modelling the actual
stellar collisions and following the early growth of the massive
runaway product. 

This paper is organised as follows. In Section~2, we
explain in more detail the scenario for forming a massive object through
runaway collisions and review the previous works on the subject. In
Section~3, we present the numerical method and physical ingredients
used in our simulations. Test simulations are presented in
Section~4. Finally, in Section~5, we summarise these results and
introduce Paper~II. In the latter, we will present the results of more than
100 simulations carried out to study runaway collisions in a variety
of clusters.

\section{THE COLLISIONAL RUNAWAY ROUTE}

\subsection{Important quantities}

Before reviewing the collisional runaway scenario, it is useful to
define some quantities which are often referred to.

If a star with mass $M_1$ and radius $R_1$ travels with relative
velocity $\Vrel$ across a field of stars of mass $M_2$ and radius
$R_2$ and number density $n_2$, the collision probability per unit
time for this star, i.e., the reciprocal of its collision time, is 
\begin{equation}
\begin{split}
\frac{1}{\tcoll^{(1,2)}} &= S_{\rm coll} \Vrel n_2 \mbox{\ \ with}\\
S_{\rm coll} &= \pi (R_1+R_2)^2
\left(1+\frac{2G(M_1+M_2)}{(R_1+R_2)(\Vrel)^2}\right).
\end{split}
\label{eq:coll_time}
\end{equation}
An important velocity scale for collisions between such stars is given
by 
\begin{equation}
\begin{split}
\Vstar^2&=2G\frac{M_1+M_2}{R_1+R_2} \\
&= \left(617.5\,\kms\right)^2 \frac{M_1+M_2}{\Msun} \frac{\Rsun}{R_1+R_2}.
\end{split}
\label{eq:vstar}
\end{equation}

In most situations, $\Vrel\ll \Vstar$ and the collision cross section
is dominated by gravitational focusing, $S_{\rm coll}\simeq 2\pi
G(M_1+M_2)(R_1+R_2)$. In a system where all stars have mass $M_\ast$,
radius $R_\ast$, density $n$ and a Maxwellian velocity distribution
with 1-D velocity dispersion $\sigmaOneD$ ($\ll \Vstar$), the collision
time (after which, each star, on average, would have experienced one
collision) is then \citep[][Eq.~8-125]{BT87}
\begin{equation}
 \tcoll \simeq 2.1\times 10^{12}\,{\rm yr}\,\frac{10^6\,\pccube}{n} 
\frac{\sigmaOneD}{30\,\kms} \frac{\Rsun}{R_\ast} \frac{\Msun}{M_\ast}\,.
\label{eq:t_coll}
\end{equation}

Two-body relaxation plays a central role in the evolution of most
clusters considered here. The local relaxation time is \citep{Spitzer87}
\begin{equation}
\begin{split}
\trlx &= 0.339 \frac{ \sigmaOneD^3 }{ G^2 \ln(\GCoulomb{\Nstar})\,n
\langle M_\ast \rangle^2 } \simeq 4.8\times 10^7\,{\rm yr}\,\times\\
&\frac{10}{\ln(\GCoulomb{\Nstar})} \left(\frac{\sigmaOneD}{30\,\kms}\right)^3 
\frac{10^6\,\pccube}{n} \left(\frac{\langle M_\ast\rangle}{\Msun}\right)^{-2}
\,,
\end{split}
\label{eq:t_rlx}
\end{equation}
where $\langle M_\ast\rangle$ is the average stellar mass and
${\Nstar}$ the total number of stars. For systems with a broad stellar
mass spectrum, we use $\GCoulomb = 0.01$ in the Coulomb logarithm (see
Sec.~\ref{subsec:nbody}). The relaxation time defined this way only
has a direct meaning for a single-mass population. In case of a mass
spectrum, it serves as a reference time but relaxational evolution may
(and does) happen on a small fraction of $\trlx$.

The core of a cluster is the central region where density and velocity
dispersion are approximately constant. We use the definition of
\citet[][Eq.\ 1-34]{Spitzer87} for the core radius,
\begin{equation}
\Rcore = \left[9{\sigmaOneD}_{,\rm c}^2/\left(4\pi G \rho_{\rm c}\right)\right]^{1/2}
\end{equation}
where $\rho=\langle M_\ast\rangle n$ and underscore 'c' indicates central
values.

\subsection{Basic scenario: summary and expectations}

\begin{figure*}
  \centerline{\resizebox{0.9\hsize}{!}{%
          \includegraphics[bb=19 159 589 701,clip]{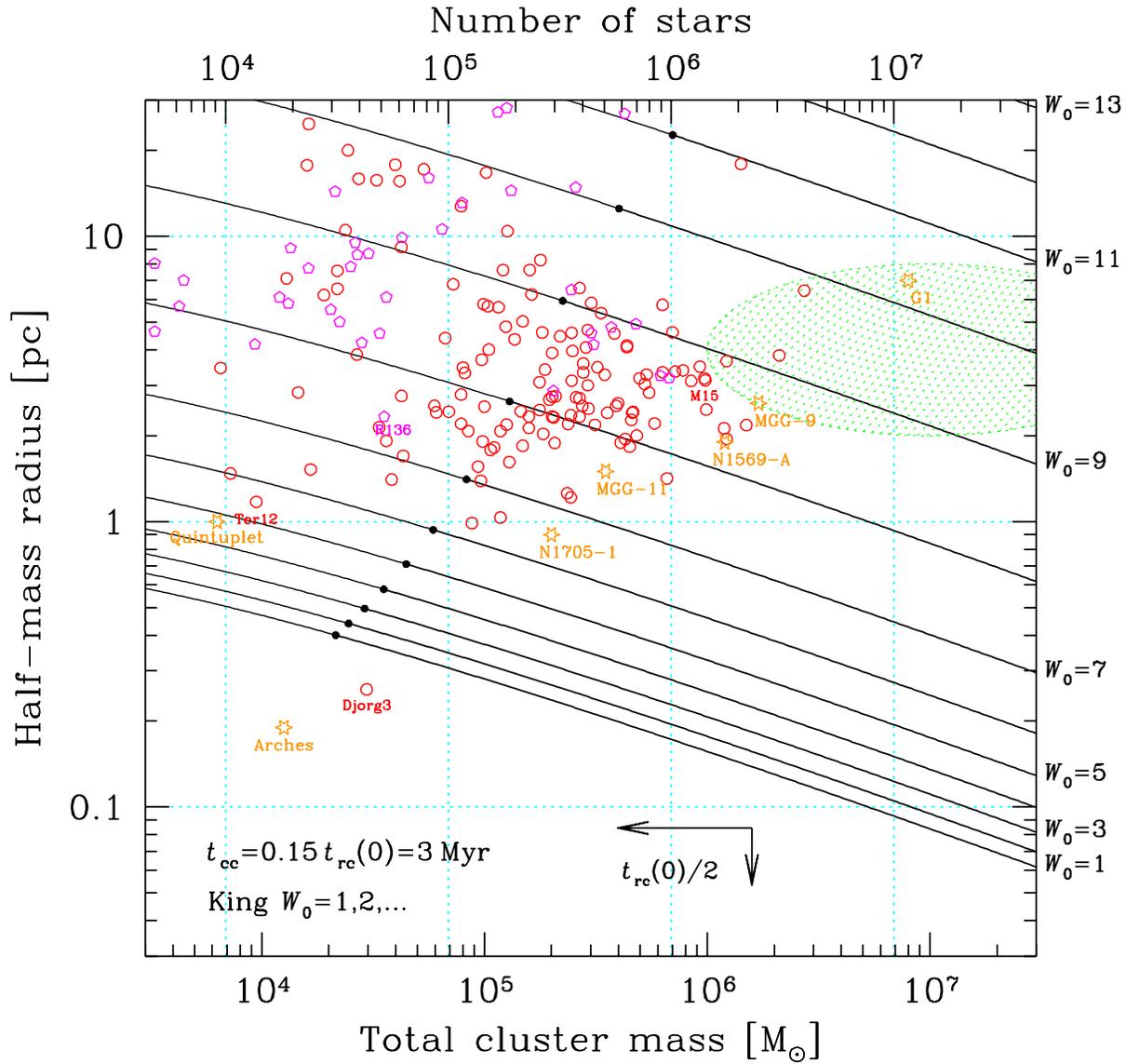}%
        }}
%
\caption{
  Conditions for rapid core collapse. This diagram shows which cluster
  masses, radii and initial concentrations will lead to core collapse
  in less than 3\,Myr, i.e., before the most massive stars evolve off
  the MS (a necessary condition for collisional runaway to occur). We
  consider clusters that have initially the structure of King models,
  with various concentrations, parametrised by the dimensionless
  central potential, $W_0$. The IMF is assumed to be Salpeter between
  $0.2$ and $120\,\Msun$ ($\left<M_\ast\right>\simeq 0.69\,\Msun$). In
  GFR04 we have showed that, for this or other similar IMF, core
  collapse occurs in $0.15\,\trc(0)$, independent of $W_0$. Each solid
  line corresponds to the condition $\tcc=3\,$Myr for a given value of
  $W_0$ (see labels on the right side of the frame). Below this line,
  the core collapse time is shorter, above, it is longer. The
  dots on the lines correspond to models with 10\,000 stars in their (initial)
  core. For this IMF, only a fraction $\sim 4\times 10^{-4}$ of the
  stars are more massive than $50\,\Msun$, corresponding of $\sim 4$
  such stars in the core. The arrows show how much of a decrease in the total
  mass $M_{\rm cl}$ or half-mass radius $R_{\rm h}$ leads to a
  shortening of $\tcc$ by a factor of 2. \newline We also show the position
  in the $(M_{\rm cl},R_{\rm h})$ plane of a variety of observed
  clusters. The circles (in red in the online color version) are the Milky Way globular clusters from
  the compilation by Harris (\citeyear{Harris96}, updated on-line at
  \url{http://physun.physics.mcmaster.ca/Globular.html}); the
  (magenta) pentagons are LMC clusters \citep{MG03}. Stars (in orange)
  represent populous young clusters,``super star clusters'' and the
  cluster G1 of M\,31. Data for the Arches and Quintuplet clusters are
  taken from \citet{Figer04}, for NGC\,1705-1 and NGC\,1569-A from
  \citet{HF96} and for MGG-9 and MGG-11 (in M\,82) from
  \citet{McCradyEtAl03}. The shaded ellipse (in green) is the region
  in which the nuclei of dwarf ellipticals and bulgeless spiral
  galaxies are observed \citep{GGvdM02,WalcherEtAl04}. For MW globular
  clusters, an age of 10\,Gyr was assumed and the total mass was
  increased to correct for the decrease of the stellar average mass
  due to evolution of massive stars (but no account has been made of
  tidal stripping of stars).}
\label{fig:FastCollPlane}
\end{figure*}

For the reader's convenience we summarise here the collisional
runaway scenario presented in GFR04 for the formation of 
an intermediate-mass black hole (IMBH, $100\,\Msun<\MBH<10^5\,\Msun$)
at the centre of a dense stellar cluster. In the next subsection
we briefly review previous works that have led to the formulation of this
scenario or have pioneered its investigation.

The basic idea is to create, through a sequence of collisions, a stellar object
much more massive than what normal star formation can produce, 
with mass of a few hundreds to a few thousands
$\Msun$. We refer to these objects as ``very massive star'' (VMS).
If such a star does not lose too much of its mass to winds by
the time it reaches the end of its life, 
it is likely to undergo complete
collapse into a black hole (BH), without mass ejection
\citep{FK01,HFWLH02}, hence producing an IMBH.

Stellar collisions are extremely rare in most astrophysical
environments. They can only play a role in systems with a high stellar
density such as self-gravitating dense clusters. When the most massive
stars initially present in a cluster explode as supernovae, the
cluster expands significantly as a result of the mass loss and
collisions stop (if they were ever occurring). Therefore the formation
of a VMS has to occur before massive stars evolve off the MS
(including the giant phase would only make a small difference), i.e.,
for a cluster containing stars up to $\sim100\,\Msun$, within a few
million years. It is possible that a cluster could be {\em born\/} in
a collisional state, i.e., with an average collision time shorter than
this, so that it would have been even more collisional while its stars
were still on the pre-MS. This has been discussed as a way of creating
all stars more massive than $\sim10\,\Msun$ \citep[][and references
therein]{BBZ98,BZ05}. However, we focus here on the simpler case of a
gas-free cluster with all stars on the MS initially.

Two-body relaxation provides a mechanism to increase the central
density of a cluster and possibly drive it to a collisional
state. Although this process of gravothermal core contraction,
eventually leading to core collapse, also operates in single-mass
clusters, it is much more important for realistic clusters with a
broad IMF. The most massive stars will then be subject to dynamical
friction and drift to the centre. In all realistic cases, the cluster
encounters the Spitzer instability, meaning that this {\em
mass-segregation} process leads to the formation of central core of
massive stars that decouples as a self-gravitating system before the
most massive stars can achieve energy equipartition with lighter
objects. The system of massive stars then experiences core collapse on
its own, very short, timescale. The high concentration of massive
stars at the centre of the cluster eventually leads to a high
collision rate.

The process of rapid mass segregation and core collapse in clusters
with a broad IMF was the subject of GFR04. There, the
relaxation-driven evolution of clusters with a variety of structures
and IMF was followed. A key finding of this work is that for a given
cluster structure but for all power-law mass spectra,
$d{\Nstar}/dM_\ast \propto M_\ast^{-\alpha}$ with $\alpha\in[1.4,3]$
as well as for Kroupa \citep{KTG93} and \citet{MillerScalo79} IMFs,
the core-collapse time $\tcc$, expressed in units of the initial
half-mass or central relaxation time ($\trh(0)$, $\trc(0)$, see GFR04)
only depends on the ratio of the maximum to the average stellar mass,
$\mu\equiv M_{\ast,\rm max}/\langle M_\ast\rangle$. Furthermore, for
$\mu>50$, a regime reached by all realistic IMF, the dependency
flattens to a constant $\tcc/\trh(0)\simeq 0.07-0.08$ for Plummer
models.
Even more interestingly, this value, when expressed in
units of $\trc(0)$ appears to be independent of the cluster's
structure\footnote{Provided one can define a non-zero $\trc(0)$, which
is not the case for $\gamma$-models with $\gamma<2$
\citep{Dehnen93,Tremaine94}.}, $\tcc/\trc(0)\simeq 0.15$. GFR04 also found
that, in contrast to core collapse in single-mass clusters, the mass
of the contracting core, instead of decreasing to zero, reaches a
finite value, representing in all cases a fraction $1.5-3\times
10^{-3}$ of the total cluster mass.

These findings led us to the following two simple expectations:
\begin{enumerate}
\item \label{item:tcc} If the dynamical role of binaries can be neglected (see
discussion in
Appendix~\ref{sec:binaries}), any cluster where the core-collapse time,
$\tcc\simeq 0.15\,\trc(0)$ (for realistically broad IMF) is shorter
than the stellar evolution time of the most massive stars present,
$\tstar\simeq\,3$\,Myr (if the mass function extends to $\sim
100\,\Msun$) will enter a phase of rapid collisions between the massive stars
that drive the core collapse. If the central velocity
dispersion at that time is not in excess of $\sim 1000 \kms$,
collisions are not too disruptive and can lead to growth by mergers
\citep{LRS93,FB05}. Since the most massive object will have the
largest cross section for further collisions, it is expected to grow
in a runaway fashion, i.e., much faster than any other star (see the
simple mathematical model in Section~\ref{subsec:previous_works}). 
\item \label{item:Mcc} When a runaway occurs, the final mass attained by the
central VMS cannot exceed about $3\times 10^{-3}$ of the total cluster
mass. This is also in agreement with the mass of the final collision
product formed in several previous $N$-body simulations of runaways
\citep{PZMMMH99,PZMcM02}.
\end{enumerate}
In paper~II, we put these expectations to the test through cluster
simulations that include stellar collisions. We shall see that, by and
large, our results confirm point \ref{item:tcc}. For large cluster
masses ($\gtrsim 10^7\,\Msun$), cluster evolution may be driven, from
the beginning, by collisions, thus accelerating collapse so that
runaway may happen at lower densities than predicted by
\ref{item:tcc}. We do not find support in our simulations for
prediction \ref{item:Mcc} but likely because our simulation method may
not be able to predict correctly the final mass attained at the end of
the runaway.

An obvious and legitimate question is whether the condition
$\tcc=\tstar=3$\,Myr is ever fulfilled in real clusters. This is
illustrated in Fig.~\ref{fig:FastCollPlane} where we show this
condition for a variety of King models in a plane representing the
(initial) mass $M_{\rm cl}(0)$ and half-mass radius $\Rh(0)$ of the
cluster.  We have assumed a $0.2-120\,\Msun$ Salpeter IMF ($\langle m
\rangle \simeq 0.69\,\Msun$, $\mu \simeq 174$).  We chose $M_{\rm
cl}(0)$ and $R_{\rm h}(0)$ as parameters because their present-day
values are more easily accessible to observation than, say, the
central density and because they probably vary less than other
quantities during cluster evolution. On the other hand, these
quantities do not by themselves determine $\trc(0)$ and, hence,
$\tcc$; one needs to know how concentrated the cluster was initially
(as measured, for instance, by the ratio of core radius to half-mass
radius), a piece of information very poorly constrained by
observations of evolved clusters. For the sake of simplicity, we
restrict our discussion to the King family of cluster models, for
which increasing $W_0$ corresponds to increasing concentration. We
note that, although most numerical simulations so far have been done
with moderate initial concentration, $W_0\le 6$, observations of young
clusters suggest they may initially have very small cores,
corresponding to $W_0\ge 8$ (\citealt{CampbellEtAl92,MDS94}; but see
\citealt{McCradyEtAl03}, concerning the core radius of R136). For
initial $W_0$ values ranging from 1 to 12, we have plotted the line
indicating clusters with $\tcc=0.15\,\trc(0)=\tstar=3\,$Myr. Below the
line, a cluster with that $W_0$ would have shorter $\tcc$. Points of
various shapes denote observed clusters (young or old; see
caption). We see that, even for a relatively moderate $W_0=7$, a few
clusters inhabit the region of parameter space for which one expects
fast core collapse, possibly leading to collisional runaway.

There is an important caveat to be made, concerning the interpretation
of the results of GFR04 and Fig.~\ref{fig:FastCollPlane}. For a
$0.2-120\,\Msun$ Salpeter IMF, only a fraction $\sim 4\times 10^{-4}$
of the stars are more massive than $50\,\Msun$. To have $10^4$ stars 
with $\sim 4$ such massive stars within the core of a $W_0=3$ or $8$
cluster, the total number of stars must be $4.2\times 10^4$ or
$1.9\times 10^5$, respectively. These values are indicated, for the
various $W_0$, as black dots on the corresponding curves of
Fig.~\ref{fig:FastCollPlane}. Actually, our result that, for
sufficiently large number of particles, the core collapse occurs on a
given fraction (0.15) of the initial {\em central} relaxation time
$\trc(0)$ does not necessarily imply that only massive stars initially
in the core (where the relaxation time is $\approx \trc(0)$) are
responsible for the process.
If it were the case, $\tcc$ would be of order
of the dynamical friction timescale in the core, $t_{\rm df,c}\simeq
\mu^{-1}\trc(0)\le 0.02\,\trc(0)$ which is is much shorter than the
value we find. This indicates that massive stars initially outside the
core have time to reach the centre.
 
\subsection{Previous works on collisional runaway}
\label{subsec:previous_works}

\citet{Colgate67} was the first to discuss the possibility of forming
stars much more massive than initially present in a dense cluster
through a sequence of stellar collisions. He suggested this mechanism
as a way to create a large population of massive stars in a galactic
nucleus to explain the quasar luminosities through enhanced super-nova
rates. He pointed out that, provided the collisions always result in
mergers with little mass loss, a runaway situation should ensue, with
one star growing to a very large mass in a short time but also
estimated that its growth would actually terminates at $\sim
50\,\Msun$ because, assuming a $R_\ast \propto M_\ast$ relation, the
runaway object would then become too diffuse to stop $1\,\Msun$
impactors (``transparency'' problem). A more realistic mass-radius
relation for MS stars more massive than $\sim30\,\Msun$ (see
Eq.~\ref{eq:MR_VMS} below) actually corresponds to nearly constant
projected mass density $M_\ast/R_\ast^2$ so that this argument does
not apply if the growing star has time to contract back to normal MS
structure between collisions.

Following \citet{Colgate67}, one may gain some insight to the runaway
mechanism by considering an idealised situation in which one star of
mass $M(t)$ and radius $R(t)$ grows by merging (without mass loss)
with stars of mass and radius $m$, $r$. One assumes $M\gg m$ (and $R
\gg r$) and a constant density $n$ of light stars, with a (3-D)
velocity dispersion $\sigmaThreeD$. Then, using
equation~(\ref{eq:coll_time}), the growth rate of the massive star reads
\begin{equation}
\begin{split}
 \frac{dM}{dt} = &\frac{m}{t_{\rm coll}} 
\simeq m n \sigmaThreeD\, \pi \left(r+R\right)^2 \left(1+\frac{2G(m+M)}{(r+R)\sigmaThreeD^2}\right) \\
 \simeq & 2\pi G \sigmaThreeD^{-1} n m M R = 
\frac{M_0}{t_0} \left(\frac{M}{M_0}\right)^{1+\beta},\\
& \mbox{with\ \ } t_0^{-1} = 2\pi G \sigmaThreeD^{-1} n m R_0.
\end{split}
\label{eq:dMdt}
\end{equation}
This holds for strong gravitational focusing and a power-law
mass-radius relation, $R = R_0 (M/M_0)^\beta$. The solution of this
differential equation, for $M(t=0)=M_0$ is
\begin{equation}
 M(t) = \frac{M_0}{\left(\frac{m}{M_0}\left(1-t/t_{\rm div}\right)\right)^{1/\beta}}
\mbox{\ \ with\ \ } t_{\rm div} = \frac{t_0}{\beta}.
\label{eq:Moft}
\end{equation}
Hence, in this toy model, $M$ becomes formally
infinite after a finite time $t_{\rm div}$, if the exponent $\beta$ is
positive. More detailed analysis of the evolution of the whole
distribution of stellar masses through use of the so-called
``coagulation equation'' also leads to the condition $\beta>0$ for
runaway to be possible \citep{LeeMH93,LeeMH00,MG02}. This kind of
approach, ignoring stellar dynamical effects as it does, can only
serve as a preliminary guide. A rough estimate of $t_{\rm div}$ in a
static cluster core would be
\begin{equation}
  t_{\rm div} = \frac{\sigmaThreeD}{2\pi \beta G n m R_0} 
\approx \beta^{-1} t_{\rm dyn} \frac{R_{\rm c}}{R_0}
\end{equation}
where $t_{\rm dyn}$ is the core dynamical time. 
One sees that this is a very long timescale because
${R_{\rm c}}/{R_0}$ is typically (much) larger than $10^6$.

\citet{Sanders70b} investigated the possibility of runaway collisions
in dense galactic nuclei (without a central (I)MBH) using a more
refined model than Colgate's. The evolution of a population of stars
subject to collisions was followed using a ``particles-in-a-box''
Monte Carlo method. The outcome of the collisions (occurrence of merger,
amount of mass and energy loss) was determined using a generalisation
of the semi-analytical method of \citet{SS66}. The structure and
dynamics of the cluster were not resolved. Instead, the system was
treated as homogeneous within a spherical domain, its size and density
being evolved by considering the amount of mass and energy lost
through evaporation of stars and collisions, respectively and assuming
permanent virial equilibrium. All the gas lost in collisions was
recycled into stars. The velocities were picked from a Maxwellian
distribution with equipartition between stars of various masses.
Thanks to a shallower $M$--$R$ relation with $\beta=0.7$, there was no
transparency saturation to the growth of a VMS, which was followed from
$0.5\,\Msun$ to more than $300\,\Msun$ in a $10^7\,\Msun$
nucleus\footnote{\citet{LS78} have stated the transparency problem in
  the following way. The growing star will be unable to stop a smaller
  impactor when the latter has more kinetic energy (at infinity) than
  what is required to punch a hole of its own cross section through
  the massive object, i.e., to unbound the mass it sweeps by passing
  through it. Neglecting the central mass concentration of the star, this
  leads to the (very approximate) condition
\[
\frac{M}{R^2}r^2\,\frac{GM}{R} \approx m \sigmaThreeD^2,
\]
where $M$ and $R$ are the mass and radius of the runaway object, $m$
and $r$ the typical values for impactors, and $\sigmaThreeD$ their
velocity dispersion. Assuming a power-law mass--radius relation
$R_\ast \propto M_\ast^\beta$ this translates into the following
relation for the ``saturation mass'' of the runaway object,
\[
 M_{\rm max} \approx m \left(\frac{v_\ast^2}{\sigmaThreeD^2}\right)^{1/(3\beta-2)}
\]
with $v_\ast^2=Gm/r$. A typical $\beta$ value for MS stars (i.e.,
assuming the runaway object stays on the MS) is $\beta\simeq 0.5$
and, except for very massive and compact clusters,
$\sigmaThreeD<0.3v_\ast\approx 300\,\kms$, so
$M_{\rm max}>100m$.}. A cluster model 10 times more massive, with the
same initial velocity dispersion of $\sim 500\,\kms$, did not exhibit
any sign of runaway because the velocity dispersion raised above
$1000\,\kms$ and thus collisions became disruptive.

\citet{LS78}, citing unpublished work by Fall and Lightman, exposed
the conditions for the onset of a collisional runaway using a simple
evaporative analytical model for the contraction of the core of a
single-mass globular cluster. They found that the core must evolve to
a state containing only a few hundreds to thousands of stars with a
velocity dispersion of order 200\,{\kms}, but such approach lacks
physical ingredients such as mass segregation which is key in
the evolution of more realistic systems. At any rate, an important
point made in this work was that, in the collisional stage, the stars
should experience mergers at such high a rate that they should have no
time to recover thermal equilibrium between two collisions and could
well stay bloated, hence bringing back the transparency problem. The
famous paper of \citet{BR78} introduced the process of runaway
collisions to a larger audience as ``one of the quickest routes to the
formation of a massive object in a dense stellar system''. However,
although they mentioned mass-segregation as a way to increase the
density of massive stars on a shorter timescale, a self-consistent
picture of the evolution a cluster subject to relaxation and
collisions was still missing.

\citet{Lee87} and \citet{QS90} were first to study the role of
collisional mergers in numerical models self-consistently resolving the
structure and evolution of stellar clusters (without a central massive BH).
They applied very similar simulation methods and assumptions to quite
different systems. Using codes that solve the Fokker Planck (FP) equation,
they had to keep the stellar mass function discretized into
``components'' and represented the cluster as a finite set of
distribution functions (in energy-space), one for each stellar mass.
This approach has the advantage of producing results virtually devoid
of noise but imposes a very artificial treatment of stellar evolution
and collisions. Stars in the same mass component have to share the
exact same properties, including some average age updated as time
passes and merger products are added to the components, a possible
cause of rejuvenation because the authors assumed complete mixing of
the stellar gas during collisions.  Collisions were treated as mergers
without any mass loss.  The mass and orbital energy of mergers of any
mass have to be cast into the predefined mass components. Another
questionable aspect of FP simulations, when applied to 
collisional runaways, is the applicability of this
formalism to mass components containing a very small number of stars
(sometimes less than one). These models
also included the dynamical formation of binaries through 3-body
interactions and their subsequent hardening (and ejection) as a
central source of energy capable of reversing core collapse and
turning off collisions in clusters with a relatively low number of
stars. 
It has since been realised that there is a high probability for
collisions to occur when these 3-body binaries interact with other
stars, leading to a very significant reduction of the heating they
provide \citep{CH96,FregeauEtAl04}.

Both \citet{Lee87} and \citet{QS90} started with clusters where all
stars initially have the same mass ($0.7$ and $1\,\Msun$,
respectively). In the work of \citet{Lee87}, who was studying globular
cluster models, collisions are actually tidal captures assumed to lead
to quick merger. He concentrated on cases for which there was no real
runaway growth or significant speed-up of core collapse due to
mass-segregation. In all models but one, the core collapse was
reversed by heating due to 3-body binaries or by mass-loss due to
stellar evolution of mergers.  As stars more massive than
$22.4\,\Msun$ were not allowed, the simulation had to be stopped for
the only case in which conditions for the runaway were met.
\citet{Lee87} suggested that a ``typical'' proto-galactic nucleus,
with $10^8$ stars and a half-mass radius of $\sim 0.4$\,pc would be
subject to the ``merger instability''. The goal of \citet{QS90} was
explicitly to look for the onset of runaway collisions as a way to
create a very massive star and, eventually, a seed IMBH that could
grow into a massive black hole (MBH, $\MBH>10^5\,\Msun$) at the centre
of a galactic nucleus, as suggested by \citet{BR78}. The results of
these simulations suggested that runaway collisions would occur
provided that the half-mass relaxation time is shorter than $\sim
10^8\,$yr (to beat stellar evolution) and ${\Nstar} \ge 3\times 10^6$
(to avoid binary heating). The authors stressed that, as a result of
mass segregation, the rise in the central velocity during collapse is
only moderate and collisions do not become disruptive. Although not
very realistic, these early studies made plausible the idea that
successive collisions and mergers of main-sequence stars could lead to
the formation of a $\sim 10^2 - 10^3\, M_\odot$ object.

Through direct $N$-body simulations of clusters containing 2000 to
65,000 stars, \citet{PZMMMH99} and \citet{PZMcM02} showed that, in
such low-${\Nstar}$ systems, dynamically formed binaries, far from
preventing collisions (by heating the cluster and reversing collapse),
actually {\em enhance\/} them by increasing the effective cross
section. In these small systems, once the few massive stars have
segregated to the centre, one of them will repeatedly form a binary
with another star and later collide with its companion when an
interaction with a third star increases the binary's eccentricity or
induces a chaotic ``resonant'' interaction. The growth of this star is
ultimately stopped by stellar evolution, or by the dissolution of the
cluster in the tidal field of the parent galaxy. Given the small
number of stars in these simulations, the maximum mass of the
collision product is only $\sim 200\,M_\sun$ when mass loss from
stellar winds is negligible.

More recently, \citet{PZBHMM04} have improved on these early $N$-body
simulations, considering models for two young clusters in the galaxy
M\,82: MGG-9 and MGG-11 \citep{McCradyEtAl03}. Most calculations for
MGG-11 were performed with 131\,072 particles (``128k''), assuming a
Salpeter IMF ranging from 1 to $100\,\Msun$; those for MGG-9 used the
same IMF and about 4 times more particles, in agreement with a higher
estimated mass. For two simulations of MGG-11, the record-breaking
number of 585\,000 particles was used in order to extend the IMF down
to a more realistic $0.2\,\Msun$ (Salpeter) and $0.1\,\Msun$
(Kroupa). The initial conditions used were King models with
dimensionless central potential ranging from $W_0=3$ to $W_0=15$. The
authors found runaway growth of a VMS in all highly concentrated
clusters ($W_0\ge 9$) with a half-mass dynamical friction timescale
for $100\,\Msun$ stars shorter than 4\,Myr. The mass reached was at
least $800\,\Msun$ and up to $2700\,\Msun$ depending in the $M$--$R$
relation. Incidentally they noted that no reasonable model for MGG-9
complies with these conditions but that MGG-11 may be in the right
domain and suggested this may explain why an ultra-luminous X-ray
source is (possibly) associated with the latter but not with the
former. In contrast to what was found in smaller systems, for models
with $>10^5$ particles (and without primordial binaries), a
significant number of collisions (of order 25--30\,\%)
\Comment{MARC}{HOLGER}{Do you confirm this figure?} involve only
single stars. \citet{PZBHMM04} also performed one (very
computer-intensive) 128k simulation for MGG-11 with $W_0=12$ and 10\,\%
primordial hard binaries and found a slight {\em increase} in the
collision rate and final mass of the runaway object, suggesting that
primordial binaries cannot prevent the runaway process by halting the
core collapse process. Whether this holds in general, in particular
for initially less concentrated clusters has yet to be established,
through, e.g., a series of MC cluster simulations including primordial binaries
(Fregeau et al., in preparation).

\section{SIMULATION METHODS AND PHYSICAL INGREDIENTS}

In the present work, we use a set of stellar dynamical simulations 
for collisional clusters to establish the
conditions and manner under which a runaway occurs. 
Our basic numerical tool is a Monte Carlo code
similar to the one used for GFR04 but developed independently and
presenting many differences in its structure. We describe this code
briefly in Section~\ref{subsec:MCcode}. The reason for using this different 
code here is that it was originally written
to study high-density galactic nuclei including the effects of
stellar collisions. In Section~\ref{subsec:Coll}, we explain our treatment
of collisions and discuss various relevant aspects of their physics.

\subsection{The Monte Carlo code for cluster dynamics}
\label{subsec:MCcode}

In the past few years, a new Monte Carlo (MC) code, {\MESSY} (for
``Monte Carlo Experiments with Spherically SYmmetric Stellar
SYstems'') has been developed to follow the long term evolution of
dense clusters, with an emphasis on galactic nuclei
\citep{Freitag01,FB01a,FB02b}. This code is based on the scheme first
proposed by \citet{Henon73} to simulate globular clusters but, in
addition to relaxation, it also includes collisions, stellar
evolution, and, optionally, tidal disruptions and captures of stars by a
central MBH through emission of gravitational waves.

The MC technique assumes that the cluster is spherically symmetric and
represents it as a set of particles, each of which may be considered
as a homogeneous spherical shell of stars sharing the same orbital and
stellar properties. Unlike with the MC code used in GFR04, in
the present implementation, the number of particles may be lower than
the number of stars in the simulated cluster (but the number of stars
per particle has to be the same for each particle). Another important
assumption is that the system is always in dynamical equilibrium so
that orbital timescales need not be resolved and the natural
time step is a fraction of the relaxation (or collision) time. Instead
of being determined by integration of its orbit, the position of a
particle (i.e., the radius $R$ of the shell) is picked up at random,
with a $R$ probability density that reflects the time spent at $R$:
$\mathrm{d}P/\mathrm{d}R\propto 1/V_\mathrm{r}(R)$ where
$V_\mathrm{r}$ is the radial velocity. Unlike the code of GFR04, our
scheme adopts time steps that are some small fraction $f$ of the {\em
  local} relaxation (or collision) time: $\delta t(R) \simeq f
\min(T_\mathrm{rel},T_\mathrm{coll})$ with $f$ of order or smaller
than 0.05\footnote{One also makes sure that the time step is
significantly shorter than any stellar evolution time (MS life-time)
for stars around $R$.}. Consequently the central parts of the cluster,
where evolution is faster, are updated much more frequently than the
outer parts. At each step, a pair of neighbouring particles is
selected randomly with probability $P_\mathrm{selec} \propto 1/\delta
t(R)$.  This ensures that a particle stays for an {\em average} time
$\delta t(R)$ at $R$ before being updated. Because particles are
evolved one pair at a time and to ensure perfect energy conservation,
the (spherical) potential produced by the collection of particles is
represented by a binary tree structure allowing both determination of
its value at any given $R$ and its update after modification of the
position or mass of a particle in $\mathcal{O}(\ln \Npart)$
operations, where ${\Npart}$ is the particle number.

The relaxation is treated as a diffusion process, with the classical
Chandrasekhar theory \citep{Chandrasekhar60,BT87}, similarly to what
is done in the code used in GFR04. This treatment shares many
assumptions with the methods that are based on integrating the 
FP equation directly. Unlike those methods, ours is based
on particles and hence allows us to incorporate further physics more naturally,
e.g., a continuous mass spectrum and the inclusion of an anisotropic
velocity distribution. The most important addition for the purposes
of this paper is stellar collisions. Unlike diffusive relaxation,
collisions cannot be treated as a continuous process but are individual
events, each of which may importantly affect the orbit, the mass or
the mere existence of a particle. When a pair is selected, one
computes the collision probability between a star of the first
particle and a star of the second,
\begin{equation}
P_\mathrm{coll} = S_\mathrm{coll} \Vrel n \, \delta t.
\label{eq:coll_prob}
\end{equation}
$S_{\rm coll}$ is given by equation~(\ref{eq:coll_time}) and is connected to
the maximum impact parameter for collision, $S_\mathrm{coll} = \pi
b_\mathrm{max}^2$. Note that the total local stellar density $n$
enters the relation, not $n_1$ or $n_2$ \citep{FB02b}.  A random
number is picked from the interval $[0,1[$ with uniform probability. If it is smaller
than $P_\mathrm{coll}$, a collision between the two particles has to
be simulated\footnote{If each particle represents $N_\mathrm{st}$
  stars, $N_\mathrm{st}$ identical collisions are assumed to take
  place. In this way, one may apply the result of the collision (new
  velocities, stellar masses,\ldots) to the particles themselves. If
  the collision results in a merger, only one particle is retained in
  the simulation. If both stars are completely disrupted, both
  particles are removed.}. The impact parameter is determined by $b =
b_\mathrm{max} \sqrt{X}$, where $X$ is another random number from the interval $[0,1[$. 
The other parameters specifying the collision, i.e., $M_1$,
$M_2$ and $\Vrel$ are already known from the particles' properties.
The method for determining the outcome of the collisions is explained in
the next section. 

\subsection{Stellar collisions and stellar evolution}
\label{subsec:Coll}

In this section, we explain some important aspects of
the ``stellar micro-physics'' of great importance for the runaway
scenario in more detail: the outcome of collisions, stellar evolution and the
interplay between them. We explain how we deal with these questions in
our code and what are the associated uncertainties.

\subsubsection{General considerations about collisions}

Consider a collision between stars of masses $M_1$, $M_2$ and radii
$R_1$, $R_2$. At large separation, their relative velocity is
$\Vrel$ and impact parameter $b$. If they were point masses, their
trajectories would be hyperbolas with separation and velocity at
periastron,
\begin{eqnarray}
d_{\rm min}&=&b\,\frac{1}{x+\sqrt{1+x^2}},\\
V_{\rm max}&=&\Vrel\left(x+\sqrt{1+x^2}\right),
\end{eqnarray}
\[
\mbox{with\ }x=\left(\frac{\Vstar}{\Vrel}\right)^2\left(\frac{R_1+R_2}{2b}\right)
\]
If one neglects tidal effects until the stars touch, the
relative velocity at contact is $V_{\rm
  cont}=\sqrt{(\Vrel)^2+\Vstar^2}$. Because of gravitational focusing,
$d_{\rm min}$ is a more useful parameter than $b$ to describe how
central a collision is. 

For $\Vrel \ll \Vstar$, gravitational focusing
is important, $S_{\rm coll}\simeq 2\pi
(R_1+R_2)G(M_11M_2)(\Vrel)^{-2}$ and $dP/d(d_{\rm min})
=\,$const, and most collisions result in merger with little mass loss,
$\delta M/M < 0.1$ \citep{BH87,BH92,LRS93,LWRSW02,SADB02,FB05}. When
$\Vrel\simeq \mbox{few}\,\Vstar$, a regime probably only reached in
galactic nuclei in the vicinity of a MBH, the cross section is
geometrical, $S_{\rm coll}\simeq \pi (R_1+R_2)^2$, most encounters
have relatively large $d_{\rm min}$ ($dP/d(d_{\rm min})
\propto d_{\rm min}$) and are ``fly-bys'', i.e., both stars survive
and remain unbound. Only nearly head-on collisions are highly
disruptive and may lead to destruction of the smaller star or both
\citep{BH87,BH92,LRS93,FB00c,FB05}.

At low relative velocities, two stars can become bound to each other,
i.e., form a binary, through dissipation of orbital energy in tides
near periastron, even if $d_{\rm min}>R_1+R_2$ \citep{FPR75}.
For velocities typical of globular clusters ($10-50\,\kms$), such
tidal binaries form up to $d_{\rm min}/(R_1+R_2)\simeq 2-3$
\citep{PZM93,KL99}. Their evolution is still a subject of debate and
may be very complex, but given the fact that they form with very small
pericentre separation and that the stars should swell due to
conversion of tidally excited oscillations into heat
\citep{MMMDT87,KG96,Podsiadlowski96}, it seems likely that they will promptly
merge. Consequently, we could try to account for tidal captures by
increasing the effective stellar radii for collisions at low $\Vrel$
\citep{Lee87}, but for simplicity and to stay on the conservative side
when testing the runaway scenario, we decided to neglect this effect
and only account for genuine collisions with $d_{\rm min}<R_1+R_2$.

\subsubsection{Mass--radius relation}

\begin{figure*}
  \resizebox{11.5cm}{!}{%
          \includegraphics[bb=29 155 570 697,clip]{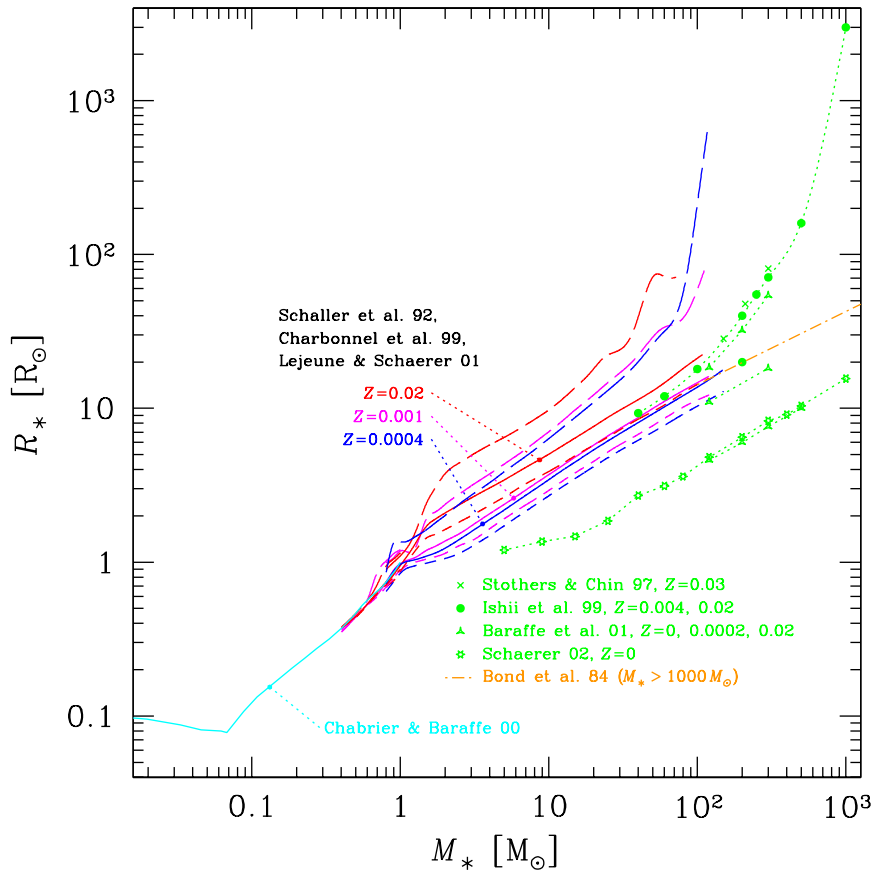}%
        }
  \hfill \parbox[b]{55mm}{%
\caption{
$M$--$R$ relations for MS stars from various
authors. The $M$--$R$ for low masses ($0.01$--$1\,\Msun$, in cyan in
the colour version) is from Fig.~3 of
\citet[][$Z=Z_\odot=0.02$, age of $5\times 10^9$ years]{CB00}
For intermediate masses, we plot radii from the Geneva stellar
evolution group for three metallicities: $Z=0.0004, 0.001, 0.02$ 
(in blue, magenta and red in the online color version of this figure)
\citep{SSMM92,CDSBMMM99,LS01}. Short-dashed lines correspond to 
the ZAMS, solid lines to the radius when the star has lived half of
its MS life-time (or $5\times 10^9$ years for low-mass stars) and
long-dashed lines to the end of the MS phase or at an age of
$10^{10}$ years at low masses. The maximum mass considered in these
series of models is 100 to $150\,\Msun$. All other $M$--$R$
relations plotted here are for the ZAMS
\citep{BAC84,SC97,IUK99,BHW01,Schaerer02}. Star with higher metallicity
have larger radii. At $Z_\odot$, stars more massive than
$100\,\Msun$ have a huge diffuse envelope, due to metal opacity.
The relation from
\citet{BAC84} ($R_\ast\simeq 1.6\,R_\odot (M_\ast/\Msun)^{0.47}$) neglects 
this effect. It is established for $M_\ast \ge 10^4\,\Msun$ but
matches nearly perfectly the models for $M_\ast\ge 100\,\Msun$.}
\label{fig:M_R}
}
\end{figure*}

One may expect the $M$--$R$ relation to play an important role in
determining collision rates through its influence on the cross
section. Fig.~\ref{fig:M_R} show various $M$--$R$ relations from the
literature. For stars between $\sim 0.8$ and $\sim 150\,\Msun$, we
plot the radius at the beginning, middle and end of the MS
\citep{SSMM92,CDSBMMM99,LS01}\footnote{Data available at
\texttt{http://obswww.unige.ch/$\sim$mowlavi/evol/
stev\_database.html}}. In the present work, the behaviour of the
$M$--$R$ relation beyond $M_\ast=100\,\Msun$ is of particular concern
because, within our assumptions, this will set the size of the runaway
collision products and may have bearing on growth and rejuvenation
rates. Hence, this may determine when stellar evolution will terminate
the growth and what mass the runaway star has attained at this
stage. Unfortunately, the radius of a VMS ($M_\ast>100\,\Msun$) is
highly uncertain. \citet{IUK99} have shown that a VMS may develop a
very extended envelope of very low density \citep[see
also][]{SC97,FNMMcLGGL98,BHW01}. Their $1000\,\Msun$, $Z=Z_\odot$
model, for instance, has $\Rstar=3000\,R_\odot$ on the zero-age main
sequence (ZAMS) but its envelope represent only 3\% of the stellar
mass so it is reasonable to assume this extended atmosphere will play
a negligible role in collisions. Furthermore, this structure is due to
opacity from metals and is not present if the metallicity is low
enough \citep{BHW01}.

For the cluster simulations presented in this paper, we adopt a unique
MS mass-radius relation constructed from $Z=10^{-3}$ models
from the following sources. For $M_\ast/\Msun<0.4$, we use Fig.~3 of
\citet{CB00}. For the ranges $0.4\le M_\ast/\Msun < 1$ and 
$1\le M_\ast/\Msun <120$ we use models from \citet{CDSBMMM99} and  
\citet{SSMM92}, respectively. Finally, for stars more massive than 
$120\,\Msun$, we apply a relation given by \citet{BAC84},
\begin{equation}
R_\ast = 1.6\,R_\odot\left(\frac{M_\ast}{\Msun}\right)^{0.47}.
\label{eq:MR_VMS}
\end{equation}
Even though this expression is for stars more massive than
$10^4\,\Msun$, it  appears to match nicely with the $M$--$R$
relation at $M_\ast/\Msun<120$, as shown in Fig.~\ref{fig:M_R}.

Focusing on low metallicity is reasonable: only if
the metallicity is sufficiently low, is it clear that a VMS will be
stable while on the MS \citep{BHW01}\footnote{But the question of the
  structure, stability and evolution of a VMS subject to a steady
  bombardment of smaller stars is another, unsettled question. Collapse
  to an IMBH may occur before the runaway collision product has evolved
  to thermal equilibrium, in which case the stability issues may
  be less relevant.}, that
it will experience little evolutionary mass loss, and that it will
collapse into a BH as a whole \citep{FK01}.

We do not account for the increase of the stellar radius during the MS
evolution and use the size of the star at half its MS lifetime, or at
an MS age of 5\,Gyr for stars with MS lifetime exceeding
10\,Gyr. Although it would be more consistent to let the stellar radii
evolve, the extra complication involved in such an improvement is not
justified in face of the larger uncertainties introduced by other
necessary simplifications, in particular concerning the size and
evolution of collision products (see below).  By the nature of the
scenario studied here, we follow only the first few million years of
the cluster evolution. Hence, low-mass stars should be given a size
smaller than their ``half-MS'' radius and closer to the ZAMS
value. One can see in Fig.~\ref{fig:M_R} that this only leads to a
slight overestimate of the collision cross sections which, being
dominated by gravitational focusing, are proportional to the stellar
radii. More important is the case of the most massive stars,
$M_\ast>50\,\Msun$, say, which are expected to dominate the
collisional process and may significantly evolve over the period of
time considered here. Their size increases significantly during the
late MS. Neglecting this may lead to an underestimate of their
collision rates. A higher, possibly more realistic, size contrast
between light and massive stars is likely to facilitate the runaway
mechanism. However, by experimenting with large changes in the
high-mass $M$--$R$ relation, ranging from $\Rstar={\rm const}$ to
$\Rstar\propto \Mstar$, we have checked that this is of relatively little
importance.

A more questionable simplification is to consider that all stars are
initially on the MS. The runaway process has to start before the most
massive stars in the IMF ($M_\ast \simeq 100\,\Msun$) turn into
compact remnants. This gives us at most 3\,Myr (see
Fig.~\ref{fig:MSdata}a). This is to be compared with the duration of
the pre-MS phase. After it has stopped accreting (and hence reached
the ``birth-line''), a pre-MS star less massive than $\sim 6-8\,\Msun$
is several times larger than on the ZAMS. The time required for
contraction onto the ZAMS strongly increases with decreasing mass; it
is shorter than 3\,Myr only for stars more massive than $2-3\,\Msun$
\citep{Palla02}. Consequently, in a young cluster, assuming for
simplicity that accretion ceased for all stars at the same instant,
low-mass stars are more extended and less dense than high-mass
objects. During this phase their collision rate is higher than in the
MS phase; however their encounters with more massive but more compact
stars may result in the tidal disruption of these low-mass objects
rather than a ``clean'' merger \citep{ZB02}. We leave these
complications out of our present study. We consider the models
presented here as the simplest possible ones that incorporate all the
key physical ingredients needed to investigate the runaway
scenario. More sophistication could be included in further works, if
deemed necessary.

\subsubsection{Collisional rejuvenation}

\begin{figure}
  \resizebox{\hsize}{!}{%
    \includegraphics[bb=71 148 485 706, clip]{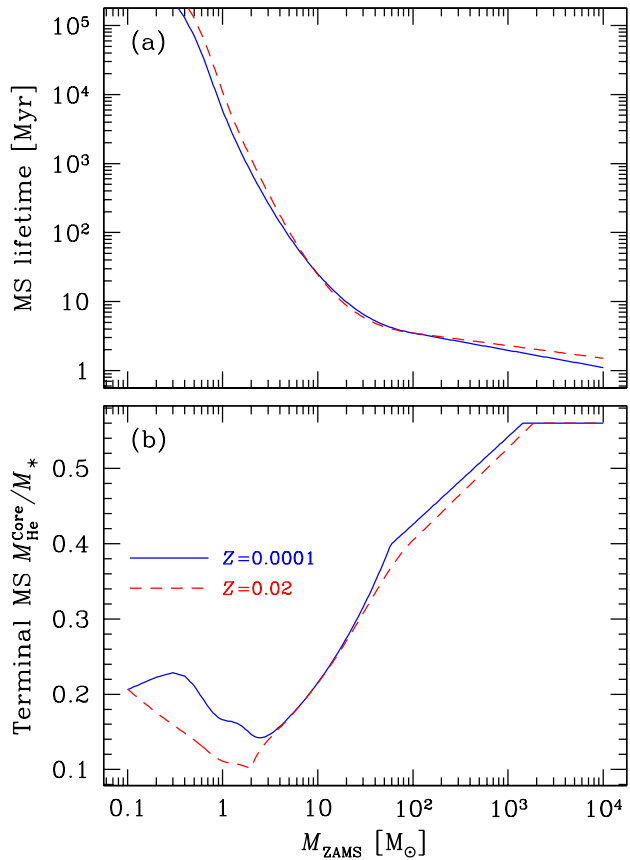}%
    }
  \caption{ (a) Main sequence lifetime for stars of various
  masses. (b) Relation between the mass of a MS star and the amount of
  helium produced in the core by hydrogen burning (at the terminal
  MS), expressed as a fraction of the total mass.  Values for two
  metallicities are plotted. Data kindly provided by
  K.~Belczynski. For masses larger than $120\,\Msun$, extrapolation
  (linear in the quantities plotted) is used; the fractional amount of
  He produced is not allowed to exceed 0.56 \citep{BAC84}.}
  \label{fig:MSdata}
\end{figure}

The stellar evolution of collision products has only been explored for
the case of low-velocity mergers, relevant to globular clusters
\citep{SLBDRS97,SFLRW00}. Such encounters are only mildly supersonic
and entropy is nearly conserved. Hence, the structure of the merger
can be established by sorting the mass elements from the parent stars
according to their entropy \citep{LRS95,LWRSW02} The main uncertainty
about the evolution of these objects is the mechanism, if any,
responsible for decreasing the amount of angular momentum as the star relaxes
to thermal equilibrium after the collision. For high velocity
collisions, significant dissipation occurs and entropy sorting is
questionable, not to mention that, in most cases, both stars survive
the collision unbound to each other (``fly-bys'').

In view of these difficulties, we used a very simple procedure to set
the stellar evolution of mergers, called {\em minimal rejuvenation.}
We assume that, during a coalescence, the helium cores of both parent
stars merge together, while the hydrogen envelopes combine to form the
new envelope; no hydrogen is brought to the core.  Furthermore, to
assign an effective age to the merger, we assume that the mass of the
helium core grows linearly with time during the MS and resort to
stellar evolution models to provide the relation between the stellar
mass and the helium core mass at the terminal age MS. This formalism
is also applied to fly-bys during which part of the stellar envelopes
is removed. In Fig.~\ref{fig:MSdata}, we plot the MS lifetime and the
total mass of helium produced during the MS phase as functions of the
stellar mass \citep[data provided by K.~Belczynski;
see][]{HPT00,BelczynskiEtAl02}. For simplicity and because the dependence on
metallicity is weak, we use the $Z=10^{-4}$ data for all simulations.
In any case, the thermal timescale is always assumed to be shorter
than the average time between collisions so that the MS mass--radius
relation is applied to collisions products. The validity of this
hypothesis during the runaway growth a massive star through repeated
mergers is questionable and is discussed in Paper~II (Section~2.2
in particular).

\subsubsection{Implementation and role of stellar evolution and mass loss}

{\MESSY} implements a very simple prescription
for stellar evolution \citep{FB02b}. While a star is on the MS, its
mass and radius are kept constant. The duration of the MS, $T_{\rm
MS}$, is given by detailed stellar models \citep{HPT00}. It is plotted
for two metallicities in panel (a) of Fig.~\ref{fig:MSdata}.  For all
simulations presented here, we use the data for metallicity
$Z=10^{-4}$ but it is obvious that the metallicity has little impact
on $T_{\rm MS}$. On the other hand, the amount of mass loss on
the MS increases strongly with $Z$ \citep{VdKL01}. While the
possibility of runaway collisions in clusters with high metallicities,
such as young populous or ``super'' clusters observed in our and other
galaxies is certainly of great interest, it is unlikely that a
high-$Z$ VMS will form an IMBH if it is left to evolve on the MS
precisely because it should experience such high mass loss. Indeed,
standard prescriptions for mass-loss rates indicate that stars more
massive than $\sim 120\,\Msun$ shed most of their mass on the MS if
they are of solar metallicity. At $Z\simeq 4\times 10^{-4}$, $\sim
120\,\Msun$ stars lose only $\sim 10$\,\% of their mass but objects
above $\sim 500\,\Msun$ should evaporate themselves nearly completely
\citep{LS01,Kudritzki02}. Therefore, in this work, we only consider
(very) low metallicity. Also recall that, in GFR04, we studied whether
mass loss on the MS could decrease the binding energy of the cluster
enough to reverse core collapse and found that, even for solar
metallicity, this only happens in a very small domain of the
parameter space, namely for clusters with a core collapse time already
very close to the critical value of 3\,Myr.

We assume that all mass lost by the stars is expelled from the
cluster. This is probably a reasonable simplification for clusters
with a relatively low escape velocity, like globular clusters but a
poor one for (proto-)galactic nuclei in which a significant fraction
of the gas could be retained. This
approximation is discussed in Paper~II.

In summary, for this work, the role of stellar evolution is only to
set a clock against which core-collapse and collisions have to race.
When core collapse requires more time than the MS lifetime of massive
stars, it will be terminated by SN mass loss (see Fig.~13 of
GFR04). This also holds for clusters of very low metallicities because
all stars with mass between $\sim 8\,\Msun$ and $\sim 40\,\Msun$ at
the end of the nuclear burning phases should explode as SNe, with
considerable mass loss \citep{FK01}. Stellar evolution of a VMS
interrupts the runaway collision sequence. If it turns into an IMBH,
it may continue to grow, although presumably at a smaller rate, by
tidally disrupting the MS stars while they still dominate the central
density.
When all normal MS stars
more massive than about $20-25\,\Msun$ have turned into stellar BH with
$M\gtrsim 10\,\Msun$, these objects will dominate the central region,
expelling MS stars and thus quenching tidal disruptions. $N$-body
simulations of systems with ${\Nstar} \le 180\,000$
\citep{BME04a,BME04b} have shown that the central IMBH will likely
form a binary with a compact remnant which, although not compact
enough to merge over a Hubble time through emission of gravitational
radiation, is probably very efficient at scattering off other stars,
thus driving cluster re-expansion and preventing any significant
growth of the IMBH. These conclusions presumably do not apply to
systems containing $10^6$ stars or more in which the binary should
have a separation small enough for quick merging (\citeauthor{BME04b}
predict $a\propto M_{\rm cl}^{-0.95}$). Recent simulations performed
with a Monte Carlo code which incorporates primordial binaries have
revealed the exciting possibility of the formation of two IMBHs. In
models with more than 10\,\% primordial binaries, a first VMS starts
growing at some distance from the centre through binary interactions
while the second forms in the centre, at the moment of core collapse
through collision between single stars, as studied here \citep{GFR06}.

The kind of MC code we use in this work does not include binary
dynamics and, more generally, is not suited for following the cluster
evolution when driven by a very small number of central objects.
Hence, we do not attempt to study the evolution of the cluster and its
central object once one VMS has formed and reached the end of its MS
phase. This would be better studied using $N$-body methods to treat
in detail the full dynamics of the central region.  

\subsubsection{SPH collision simulations}
\label{Subsec:SPHColl}

In the MC method, collisions are handled on an event-by-event basis,
rather than through average rates and outcomes as in Fokker-Planck
codes. To treat collisions with as much realism as possible, we
can use the results of 3-D hydro simulations performed with the Smoothed
Particle Hydrodynamics (SPH) method \citep{Benz90,Monaghan92,RL99}.
A set of some 14\,000 SPH simulations of collisions between MS stars
was performed by
\citet{FB05}. The original goal was to include the effects of collisions 
in stellar dynamical models of galactic nuclei \citep{FB02b}. This
focus determined the parameter range covered in this study: stellar
masses from $M_{\rm min}=0.1\,\Msun$ to $M_{\rm max}=74.3\,\Msun$,
relative velocities in the range $\Vrel/\Vstar \simeq 0.03$--$30$ and
impact parameter corresponding to $d_{\rm
min}/(R_1+R_2)=0$--$0.9$. Because it proved intractable to summarise
the outcome of these simulations (stellar masses, orbital energy and
deflection angle) through a set of fitting formulae, we implemented a
scheme to interpolate from SPH results in the four-dimensional
parameter space $(M_1,\,M_2,\,\Vrel,\,d_{\rm min})$. While this method
proves adequate for galactic nuclei simulation in which velocities are
high and it is therefore highly unlikely that mergers will lead to
formation of a star more massive than $M_{\rm max}$
\citep{TheseFreitag,FGR04c}, it is not suited to the problem of
runaway growth.

In most cluster simulations reported here and in Paper~II, we simply
assume that all collisions result in merger with no mass-loss. We will
see that this is a satisfying approximation because, in all but the
most extreme cases, the central velocity dispersion is and remains
relatively small. To go one step beyond this zeroth order ``sticky
sphere'' approximation and test its validity, for a subset of runs, we
use the SPH results to allow for non-merging collisions and
collisional mass loss in the way described below. For grazing
collisions not resulting in a merger, we do not account for the
reduction of the relative velocity or non-Keplerian deflection angle.

Complete disruption of both stars is extremely unlikely. Not only does
it require a relative velocity of a few $\Vstar$ but also a nearly
head-on geometry. Hence, in this work, we consider only two possible
outcomes: merger and ``fly-by''. Most low-velocity
($\Vrel<0.1\,\Vstar$, say) encounters result in mergers but, in our
SPH work, we have not studied those in much detail. Such collisions
require a large amount of computation time because, unless the
collision is nearly head-on, the pair does not merge immediately after
first contact but forms a bound binary which goes through many
successive periastron passages until final coalescence. In most cases,
the hydrodynamical simulation was stopped before a merged star had
formed. But it is very likely that any binary formed through a contact
collision will eventually merge because the stars swell as their
envelopes are shocked so that each pericentre passage is more
dissipative. Hence, we can use our SPH results to establish the
condition for merger. We have found the following parameterisation to
correctly predict the maximum merger impact parameter for all but a
few SPH simulations:
%
%
%
\begin{equation}
\begin{split}
\lambda_{\rm merg} & = c_0 + c_1 q + c_2 M_2 + c_3 l_v + c_4 l_v q + c_5 l_v M_2 + \\ 
& c_6 l_v^2 + c_7 l_v^2 q + c_8 l_v^2 M_2, \\
%
\mbox{with\ \ }
\lambda & =\frac{d_{\rm min}}{R_1^{({\rm h})}+R_2^{({\rm h})}},\ \ q = M_1/M_2 \le 1\\
\mbox{and\ \ }
l_v & = \log_{10} \frac{{\Vrel}}{\Vstar^{({\rm h})}},\ \ 
\Vstar^{({\rm h})}=\sqrt{\frac{2G(M_1+M_2)}{R_1^{({\rm h})}+R_2^{({\rm h})}}}.
\label{eq:merg_criterion}
\end{split}
\end{equation}
Index 1 indicates the less massive star; $R_{1,2}^{({\rm h})}$ are the
radii enclosing half the mass of each star. The numerical coefficient
are $( c_0,\,c_1,\ldots\, c_8 ) = ( 0.525$, $-0.107$, $6.84\times10^{-3}$,
$-2.03$, $0.525$, $6.16\times10^{-4}$, $0.132$, $0.526$, $-4.89\times10^{-3}
)$.

To determine the final stellar masses, we make use of the SPH mass
loss results in the following way. First when $M_2>M_{\rm
max}=74.3\,\Msun$, we try to rescale $M_1$ and $M_2$ by some factor
$\eta$ bringing $\eta M_2$ to $0.95\,M_{\rm max}$ while keeping $\eta
M_1> M_{\rm min}$. This would conserve the mass ratio $q$ but is not
possible when $q<q_{\rm min}=M_{\rm min}/M_{\rm max}\simeq
10^{-3}$. Such cases are relatively unimportant because, in typical
cases, the VMS grows mostly by merging with $50-120\,\Msun$ stars so
$q>0.025$ even for $M_{\rm VMS}=2000\,\Msun$. We deal with them by
rescaling $M_1$ to $1.05\,M_{\rm min}$, independently of $M_2$. From
the SPH results, we interpolate the fractional mass loss $\delta_M=(\delta
M_1+\delta M_2)/(M_1+M_2)$ for the $(M_1,\,M_2,\,\Vrel,\,d_{\rm min})$
parameters of the collision (with $M_1$, $M_2$ possibly rescaled), as
explained in \citet{FB05}. Inspection of SPH results for fly-bys shows
that only in very exceptional cases does the mass of any of the two
stars increase. In fact, to a good approximation, we find the mass
ratio to be conserved. We thus use this assumption to determine the
individual masses from $\delta_M$.

\section{TEST SIMULATIONS}

Before embarking on the large-scale numerical exploration of the
runaway scenario to which Paper~II is devoted, we first demonstrate
that our numerical tools are up to this task. Here we check that we
can reliably model the following two key aspects of the scenario: (1)
Fast segregation-driven core collapse in clusters with a broad mass
function and (2) Influence of collisions in cluster evolution and
onset of collisional runaway.

\subsection{Important quantities and units}
\label{subsec:inicond_units}

\begin{table}
  \caption{
Important quantities for the Plummer model. All quantities are in $N$-body units except times which are in FP units.}
    \setlength\tabcolsep{3pt}
    \begin{center}
  \begin{tabular}{lll}
    \hline
    Quantity                  & Symbol         & Value \\
 \hline
    Plummer scale             & $R_{\rm P}$    & $3\,\pi/16 \simeq 0.589$ \\
    Core radius               & $\Rcore$       & $0.417$ \\
    Half-mass radius          & $\Rh$          & $0.769$ \\
    Central density           & $\rho_{\rm c}$ & $1.167$ \\
    Central 1D velocity dispersion & $\sigma_{v,\rm c}$ & $0.532$ \\
    Mass within $\Rcore$      & $\Mcore$       & $0.193$ \\
    Central relaxation time   & $\trc$         & $0.0437$ \\
    Half-mass relaxation time & $\trh$         & $0.0930$ \\
    \hline
  \end{tabular}
\end{center}
  \label{table:Plummer}
\end{table}

Keeping with the tradition, when not stated otherwise, we are using
the $N$-body unit system \citep{Henon71a} defined by $G=1$, $M_{\rm
cl}(0)=1$ (initial total cluster mass) and $U_{\rm cl}(0)=-1/2$
(initial cluster potential energy). As time unit, we prefer the
``Fokker-Planck'' time $T_{\rm FP}$ to the $N$-body unit $T_{\rm NB}$
because the former is a relaxation time while the latter is a
dynamical time; they are related to each other by $T_{\rm FP} =
\Nstar/\ln(\GCoulomb\Nstar)\, T_{\rm NB}$. 

All test computations performed here are based on the Plummer model
\citep{Plummer11,BT87} for which we recall some important quantities
in Table~\ref{table:Plummer}.

We refer to GFR04 (Section~3 and Table 1) for more detailed
explanations about units and the important physical parameters of a
variety of clusters models.

\subsection{Core collapse without collisions}

\subsubsection{Comparison with direct $N$-body simulations}
\label{subsec:nbody}

\begin{figure}
  \resizebox{\hsize}{!}{%
          \includegraphics[bb=20 148 573 689,clip]{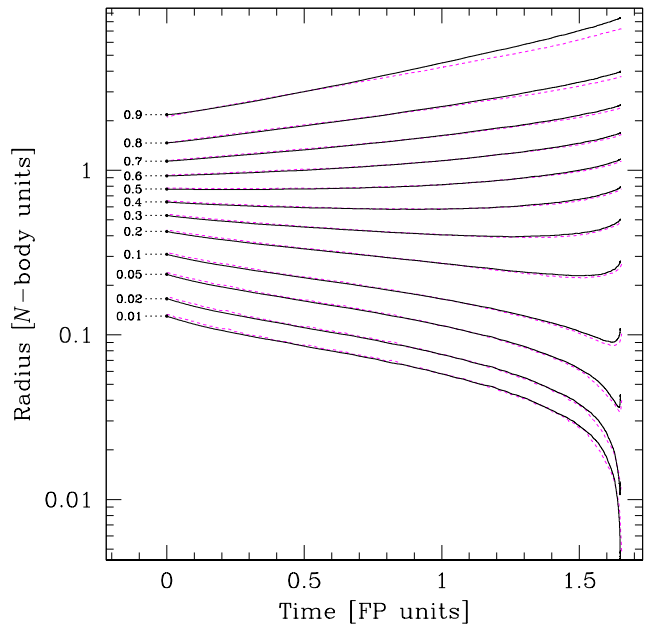}%
        }
\caption{
Core collapse of a single-mass Plummer model. We compare the results
of {\MESSY} with $\Npart=300\,000$ particles (solid lines) to those of
a direct $N$-body run with $\Npart=65\,536$ (dotted lines, in magenta
in the on-line colour version; from \citealt{BHHM03}). We plot the
evolution of various Lagrange radii, i.e., radii of spheres enclosing
the indicated fraction of the cluster mass. For the $N$-body
simulation, only fractions larger or equal to 1\,\% are
tracked. Fractions of 3, 4,\ldots 9\,\% are tracked in the $N$-body
run but not in the MC computation. A value of $\GCoulomb=0.09$ was
used in the Coulomb logarithm when converting $N$-body time units into
FP units to get the best agreement in core collapse times. To improve
clarity, the curves have been smoothed using a sliding average with
Gaussian kernel.}
\label{fig:comp_Nb_single_R}
\end{figure}

\begin{figure}
  \resizebox{\hsize}{!}{
          \includegraphics[bb=20 148 581 689,clip]{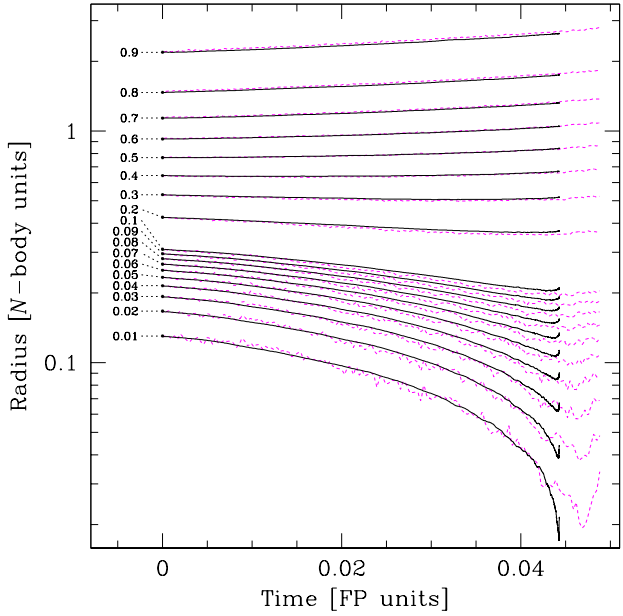}%
        }
\caption{
Evolution of Lagrange radii during the core collapse of a Plummer
model with a mass spectrum. The cluster has a Kroupa mass function
(see text) extending from 0.1 to $10\,\Msun$. We compare the results
of {\MESSY} with $\Npart=10^6$ particles (solid lines) to those of
a direct $N$-body run with $\Npart=131\,072$ (dotted lines, in magenta
in the on-line colour version). A value of $\GCoulomb=0.015$ was used
in the Coulomb logarithm when converting $N$-body time units into FP
units to get the best overall agreement.}
\label{fig:comp_Nb_10_R}
\end{figure}

\begin{figure}
  \resizebox{\hsize}{!}{%
          \includegraphics[bb=30 151 563 689,clip]{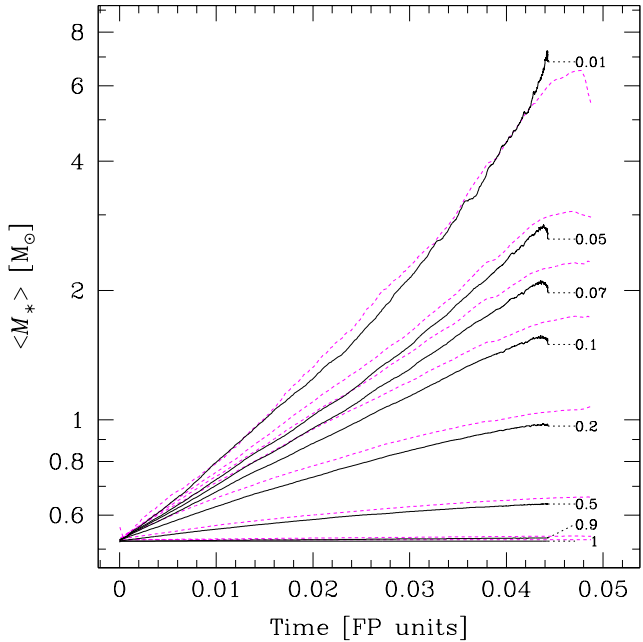}%
        }
\caption{
Mass segregation during the core collapse of a
Plummer model with a mass spectrum. The data is for the same MC and
$N$-body simulations as in Fig.~\ref{fig:comp_Nb_10_R}. We plot the
stellar mass averaged over all particles inside spheres containing the
indicated fraction of the total mass of the cluster.}
\label{fig:comp_Nb_10_M}
\end{figure}

\begin{figure}
  \resizebox{\hsize}{!}{
          \includegraphics[bb=20 148 581 689,clip]{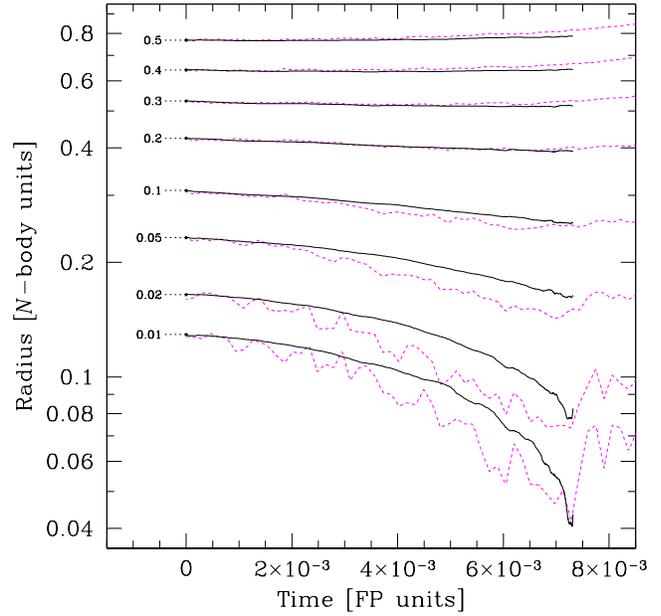}%
        }
\caption{
Evolution of Lagrange radii during the core collapse of a Plummer
model with a broad mass spectrum. The cluster has a Kroupa mass
function (see text) extending from 0.1 to $100\,\Msun$. We compare the
results of {\MESSY} with $\Npart=1.25\times 10^6$ particles (solid
lines) to the averaged results of two direct $N$-body runs with
$\Npart=131\,072$ (dotted lines, in magenta in the on-line colour
version). Our standard value of $\GCoulomb=0.01$ was used in the
Coulomb logarithm when converting $N$-body time units into FP units.
}
\label{fig:comp_Nb_100_R}
\end{figure}

\begin{figure}
  \resizebox{\hsize}{!}{%
          \includegraphics[bb=30 151 563 689,clip]{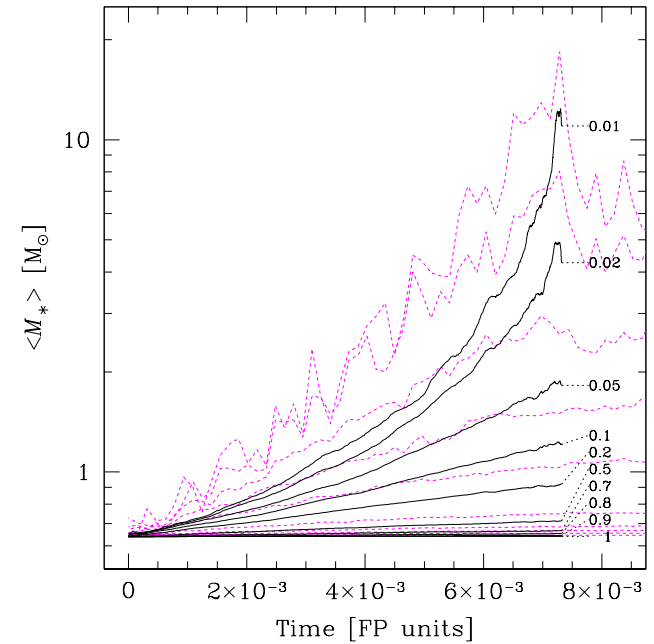}%
        }
\caption{
Mass segregation during the core collapse of a
Plummer model with a broad mass spectrum. The data is for the same MC
and $N$-body simulations as in Fig.~\ref{fig:comp_Nb_100_R}. See
caption of Fig.~\ref{fig:comp_Nb_10_M} for explanations about the
plotted quantities.}
\label{fig:comp_Nb_100_M}
\end{figure}

In \citet{FB01a}, the ability of {\MESSY} to follow the relaxational
evolution of clusters has been successfully tested against other
simulation methods for clusters with single-mass, 2-component or
continuous mass functions. However, in that paper, no comparison was
done with direct $N$-body results and the only continuous mass
spectrum considered was a relatively narrow $0.1-1.5\,\Msun$ Salpeter
IMF ($\mu\equiv M_{\ast,\rm max}/\langle M_\ast\rangle \simeq
2.2$). Given the importance of core collapse in clusters with
$\mu> 100$ for the runaway scenario and the dearth of published data
about this situation, we decided to carry out new comparisons between
{\MESSY} and direct $N$-body integrations of the core-collapse
evolution of clusters with various IMFs.

The $N$-body simulations reported here were done by H.~B. with
the collisional Aarseth $N$-body code {\NBFOUR} \citep{Aarseth99} on the
GRAPE-6 boards of Tokyo University \citep{MFKN03}. Use of GRAPE-6
hardware is essential to perform $N$-body simulations with more than
$10^5$ stars within a reasonable amount of computer time. Details on
the {\NBFOUR} code can be found in \citet{Aarseth99} and references
therein and \citet{BM03}.

We first checked the simplest situation, that of a single-mass
cluster. In Fig.~\ref{fig:comp_Nb_single_R}, we compare the evolution
of the Lagrange radii of a Plummer model as obtained with {\NBFOUR}
\citep{BHHM03} and {\MESSY}. The agreement between the two methods is
excellent if we set the coefficient in the Coulomb logarithm to
$\GCoulomb=0.09$ when converting the $N$-body time units, natural to
{\NBFOUR}, to FP units. Given the run-to-run variations of MC results
(for different random sequences), this value is compatible with the
one found by
\citet{GH94a} in comparisons between $N$-body runs with various particle numbers
($\Npart=250-2000$) and later confirmed by \citet{Baumgardt01}
($\Npart=128-16\,384$), $\GCoulomb=0.11$; it is also compatible with
the one chosen by \citet{DCLY99} to adjust their FP results to
$N$-body simulations, $\GCoulomb=0.10$.

After this test-run, we considered a cluster consisting of stars with
masses from $0.1$ to $10\,\Msun$. The masses are distributed according
to the IMF advocated by \citet{Kroupa00b}. For our simulations, the
``Kroupa IMF'' corresponds to a piecewise power law:
$d\Nstar/d\Mstar\propto \Mstar^{-\alpha}$ with $\alpha=1.3$ below
$0.5\,\Msun$ and $\alpha=2.3$ for higher masses. This produces a mass
ratio $\mu\simeq 19.3$. Fig.~\ref{fig:comp_Nb_10_R} depicts the
Lagrange radii evolution for {\MESSY} and {\NBFOUR} simulations of
this model. Converting time units with $\GCoulomb=0.015$, we observe
an excellent agreement between the results of both codes until, at
$t\simeq 0.043\,T_{\rm FP}$, shortly before the moment of deepest
collapse in the $N$-body run, it starts showing slower contraction of
the innermost regions. This departure from the MC results is probably
due to small-number effects, such as large-angle scatterings or binary
formation, that naturally kick in in the shrinking core of the
$N$-body system (computed with $\Npart=131\,072$) but are, by nature
of the MC approach, absent from the {\MESSY} run. Eventually, at
$t\simeq 0.047\,T_{\rm FP}$, 3-body binaries reverse core collapse in
the $N$-body simulation. \citet{Henon75} explained why a smaller
$\GCoulomb$ value is appropriate for systems with a mass spectrum and,
from multi-mass $N$-body simulations with $\Npart=250-1000$,
\cite{GH96} indeed found $\GCoulomb\simeq 0.015-0.025$.

In such multi-mass clusters, core collapse is
driven by mass segregation. Fig.~\ref{fig:comp_Nb_10_R} allows us to
witness this process by plotting the evolution of the average mass
within spheres containing various fractions of the total cluster
mass. Again the MC results match those of the $N$-body run closely.

We now consider a mass spectrum of realistic breadth, i.e., a Kroupa
IMF extending from $0.1$ to $100\,\Msun$, yielding $\mu \simeq
157$. To reduce numerical noise without increasing $\Npart$ to an
impractically high value, we realised two $N$-body simulations of this
system with $\Npart=131\,072$ (starting from different realisations
of the initial cluster) and averaged their results. Comparison of the
Lagrange radii and average stellar mass evolutions are shown in
Figs~\ref{fig:comp_Nb_100_R} and \ref{fig:comp_Nb_100_M}. This
time, we stuck to $\GCoulomb=0.01$, the value we traditionally use to
convert from FP time to physical time in our MC simulations of
multi-mass clusters. This value turns out to yield a perfect
match between the MC core-collapse time and the time of maximum
contraction of the inner regions in the $N$-body runs. However,
$\GCoulomb\simeq 0.025$ would have allowed a better agreement in the
early shapes of the Lagrange radii curves. At any rate, the agreement
between {\MESSY} and {\NBFOUR} results is not as good as for previous
cases. In particular, it appears that mass segregation is faster and
stronger but more progressive in the $N$-body run. In the MC
simulation, segregation accelerates at late times and the average
stellar mass in the innermost regions (inside the 1\,\% Lagrange
radius) reaches values similar to those found in the $N$-body run near
the moment of collapse. A better understanding of the cause of the
differences between the two simulation methods in the regime of broad
IMF may be reached in future investigations thanks, in particular, to
$N$-body simulations with higher $\Npart$ and, hence, less noise and
less affected by small-number effects in the central regions. For the
moment, we note that the {\MESSY} evolution of this broad-IMF cluster is
qualitatively similar to that shown by the direct $N$-body integration
and that good quantitative agreement is obtained for the aspects most
important to the present investigation of the collisional runaway,
namely the core collapse time and the degree of mass segregation
reached in the central regions during late collapse. The core collapse
time we obtain with {\MESSY} is $\tcc\simeq 7.3\times 10^{-3}\,T_{\rm
FP} = 7.9\times 10^{-2}\,\trh(0) = 0.17\,\trc(0)$, in agreement
with the value of $\tcc\simeq 0.15\,\trc(0)$ found in GFR04 for
systems with $\mu \gtrsim 50$.

As for the value of $\GCoulomb$, we decided to stay on the
conservative side by keeping it at 0.01. This is probably slightly too
small, hence predicting a core-collapse evolution too {\it
slow} with respect to other time scales, most importantly that for
stellar evolution, but the difference in relaxation time compared to,
say, $\GCoulomb\simeq 0.025$ is smaller than 10\,\% for
$\Nstar\ge10^6$.

\subsubsection{Comparison with the gaseous model}
\label{subsec:spedi}

\begin{figure}
  \resizebox{\hsize}{!}{
	\includegraphics[bb=22 148 573
  695,clip]{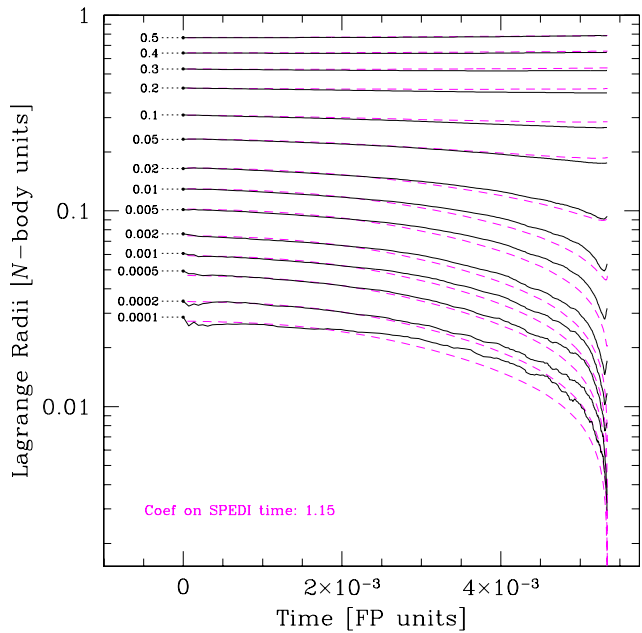}
}
\caption{
Evolution of Lagrange radii during the core collapse of a Plummer
model with a broad mass spectrum. The cluster has a Salpeter mass
function extending from 0.2 to $120\,\Msun$. We compare the results of
{\MESSY} with $\Npart=1.25\times 10^6$ particles (solid lines) to
those obtained with {\SPEDI} (with dotted lines, in magenta in the
on-line colour version). In order to allow a better comparison of the
shapes of the curves, we aligned the core collapse times by multiplying
the time units for the {\SPEDI} run by a factor 1.15.}
\label{fig:comp_SPEDI_R}
\end{figure}

The direct $N$-body method represents the most accurate but most
computationally expensive way of simulating the secular evolution of a
stellar cluster, subject to relaxation. Unfortunately, precisely
because it treats gravitation in such a direct fashion, it offers no
or little clean and easy way of establishing the global behaviour that
a system may exhibit in the limit of very large number of stars
($\Nstar\gg 10^5$) from the results of simulations made with much
fewer particles and ridden with noise and small-$N$ effects.

A nearly opposite approach is taken in {\SPEDI}, a cluster evolution
code developed by R.~Spurzem and collaborators
\citep{LS91,GS94,ST95,ASFS04}\footnote{{\SPEDI} is maintained at 
the Astronomisches Rechen-Institut in Heidelberg, see
\texttt{http://www.ari.uni-heidelberg.de/ gaseous-model}.}. Here, like
in the MC scheme, the star system is treated in a explicitly
statistical manner which assumes a very large number of stars. By
taking velocity moments of the Boltzmann equation up to the second
order, one obtains a set of partial differential equations for the
evolution of density, average (radial) velocity and velocity
dispersions (radial and tangential) of the stars at each position in
the (spherical) cluster. These equations are similar to those
governing the structure of self-gravitating spherical gas cloud. The
effects of 2-body relaxation are treated through local prescriptions
for heat conduction among stars of the same mass and energy exchange
between stars of different masses. This method, although
approximative, allows fast simulations and smooth results because the
cluster is treated as a continuum. In this scheme, the mass spectrum
has to be discretized into a number of components, each representing
stars of a given mass.

In Fig.~\ref{fig:comp_SPEDI_R}, we compare the evolution of Lagrange
radii, as obtained with {\MESSY} and {\SPEDI}, for a cluster with a
Salpeter IMF ($d\Nstar/d\Mstar\propto \Mstar^{-2.35}$) extending from
$0.2$ to $120\,\Msun$. The {\SPEDI} simulation is the same as
presented in Fig.~2 of GFR04; it was carried out using $N_{\rm
comp}=50$ mass components. Our tests show that the results depend very
little on $N_{\rm comp}$ as soon as more than $\sim 12$ components are
used. We have set the gaseous model parameters to their standard
values: the coefficient in the thermal conductivity relation is
$\lambda=0.4977$ \citep{GS94} and equipartition time is set by
$\lambda_{\rm eq}=1$ \citep{ST95}. We note that {\SPEDI} produces a
collapse some 13\,\% faster than {\MESSY}, a difference that may be
the result of an slightly inadequate value of $\lambda_{\rm eq}$ as
this parameter has only be determined for the case of 2-component
models, by comparison with $N$-body and FP simulations
\citep{ST95}. In the figure, we have rescaled the time of the 
gaseous-model simulation to obtain the same $\tcc$ as in the MC
run. The evolution to core collapse is just slightly more progressive
in the {\SPEDI} run but the agreement is otherwise very good, given
the important differences in both numerical approaches. The core
collapse time we obtain is $\tcc\simeq 5.4\times 10^{-3}\,T_{\rm FP} =
5.8\times 10^{-2}\,\trh(0) = 0.12\,\trc(0)$, $\sim 20$\,\% faster than
in GFR04. This difference reflects not only differences between the
two MC codes but also intrinsic variations in $\tcc$ obtained between
{\MESSY} runs computed for the same initial conditions but with different
random sequences (see Fig.~2 of Paper~II).

\subsection{Runaway collisions: Comparison with Quinlan \& Shapiro (1990)}

\begin{table*}
  \caption{Cluster simulations for comparison with \citet{QS90}.}
    \setlength\tabcolsep{3pt}
    \begin{center}
  \begin{tabular}{lclllll}
    \hline
    Name                        & Name in QS90 & \multicolumn{1}{c}{$\Nstar$}&\multicolumn{1}{c}{$\Npart$}&\multicolumn{1}{c}{$\Rnb$}&\multicolumn{1}{c}{$\tcc$}        &\multicolumn{1}{c}{$\tcc$}\\
                                &              &                             &                            &\multicolumn{1}{c}{(pc)}  &\multicolumn{1}{c}{($T_{\rm FP}$)}&\multicolumn{1}{c}{(Myr)}\\
 \hline
    \Sim{QS90-E4A}             & E4A          & $1.8\times 10^7$ & $5\times 10^5$   & $0.41$  & $0.134$        & $141$    \\
    \Sim{QS90-E4Ahr}           & E4A          & $1.8\times 10^7$ & $2.1\times 10^6$ & $0.41$  & $0.129$        & $136$    \\
    \Sim{QS90-E4Asph}$^{\rm (a)}$& E4A          & $1.8\times 10^7$ & $5\times 10^5$ & $0.41$  & $0.275$        & $289$    \\
    \Sim{QS90-E4B}             & E4B          & $3.1\times 10^7$ & $5\times 10^5$ & $0.71$  & $0.182$        & $553$    \\
    \Sim{QS90-E4Bb}$^{\rm (b)}$& E4B          & $3.1\times 10^7$ & $5\times 10^5$ & $0.71$  & $0.172$        & $522$    \\
    \Sim{QS90-E4Ac}$^{\rm (c)}$& E4B          & $3.1\times 10^7$ & $5\times 10^5$ & $0.71$  & $0.189$        & $574$    \\
    \Sim{QS90-E2A}             & E2A          & $6.2\times 10^6$ & $5\times 10^5$ & $0.565$ & $0.371$        & $397$    \\
    \Sim{QS90-E2B}             & E2B          & $1.1\times 10^7$ & $5\times 10^5$ & $1.00$  & $^{\rm (d)}$   & $\cdots$ \\
    \Sim{QS90-E1A}             & E1A          & $2.1\times 10^6$ & $5\times 10^5$ & $0.765$ & $0.758$        & $803$    \\
    \Sim{QS90-E1B}             & E1B          & $3.6\times 10^6$ & $5\times 10^5$ & $1.31$  & $1.23$         & $3680$   \\
    \Sim{QS90-E4Ab}$^{\rm (e)}$& E1B          & $3.6\times 10^6$ & $5\times 10^5$ & $1.31$  & $1.24$         & $3710$   \\
    \hline
    \multicolumn{7}{l}{
      $^{\rm (a)}$Collision prescriptions based on SPH data. Minimal rejuvenation.}\\
    \multicolumn{7}{l}{
      $^{\rm (b)}$Other random sequence than for \Sim{QS90-E4B}.}\\
    \multicolumn{7}{l}{
      $^{\rm (c)}$Time step four times larger than for \Sim{QS90-E4B}.}\\
    \multicolumn{7}{l}{
      $^{\rm (d)}$Core collapse stopped by stellar evolution.}\\
    \multicolumn{7}{l}{
      $^{\rm (e)}$Other random sequence than for \Sim{QS90-E1B}.}\\
  \end{tabular}
\end{center}
  \label{table:QS90}
\end{table*}

\begin{figure}
  \resizebox{\hsize}{!}{ \includegraphics[bb=23 154 582
  696,clip]{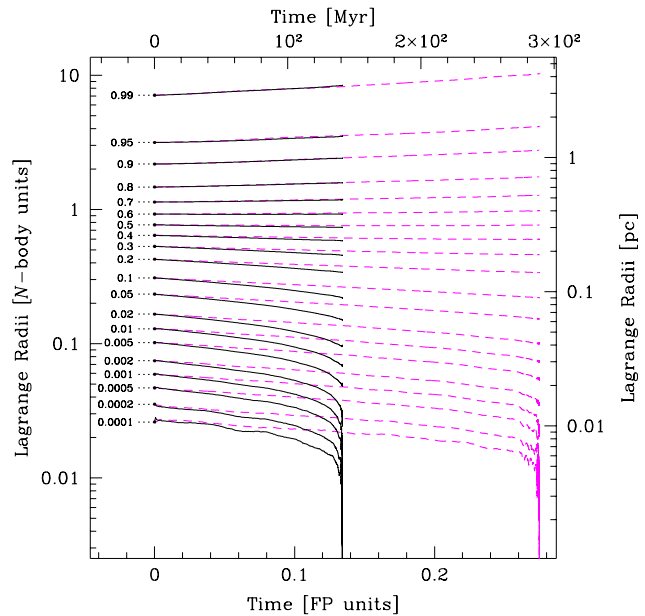} }
\caption{Evolution of Lagrange radii in core collapse with stellar collisions. 
The initial conditions are those of model E4A of QS90. We present two
simulations. In the first one (solid lines, \Sim{QS90-E4A}), all
collisions are treated as pure mergers with complete rejuvenation. In
the second case (dashed lines, \Sim{QS90-E4Asph}), we use SPH-inspired
collision prescriptions and minimal rejuvenation.}
\label{fig:QS90E4A_LargRad}
\end{figure}

\begin{figure}
  \resizebox{\hsize}{!}{
          \includegraphics[bb=32 148 563 690,clip]{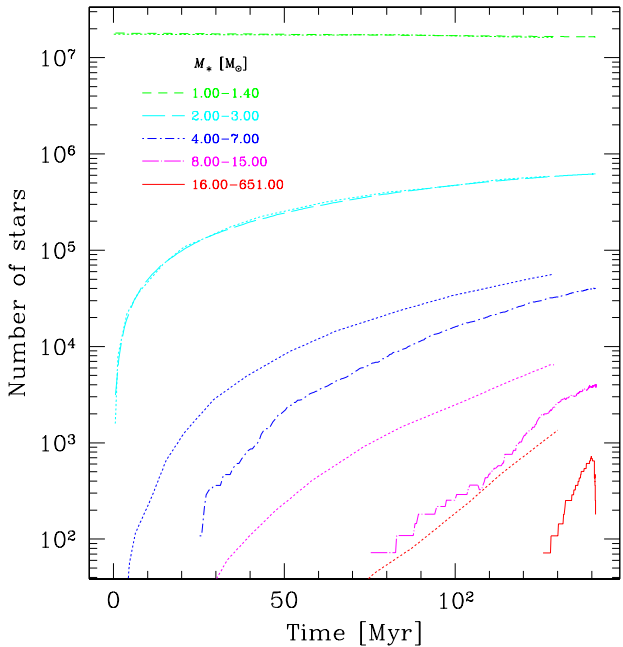}%
        }
\caption{
Number of stars in various mass bins during cluster evolution with
stellar collisions. Same simulation as in
Fig.~\ref{fig:QS90E4A_LargRad} (\Sim{QS90-E4A}). Dotted lines are the
results of QS90 (their Fig.~2a) for mass-component of mass 1, 2, 4, 8
and $\ge 16\,\Msun$. The curves for 1 and $2\,\Msun$ are nearly
indistinguishable from ours (for $1-1.4$ and $2-3\,\Msun$ bins). We
obtain a lower number of stars more massive than $3\,\Msun$ than
QS90. At the end of our simulation, $\sim 200$ white dwarfs and a
similar number of neutron stars have formed. The (negligible)
contribution of neutron stars is included in the first mass bin; white
dwarfs are not plotted. We used $5\times 10^5$ particles for our
simulation so each of them represents 36 stars.}
\label{fig:QS90E4A_Nstars}
\end{figure}

\begin{figure}
  \resizebox{\hsize}{!}{
          \includegraphics[bb=32 148 563 690,clip]{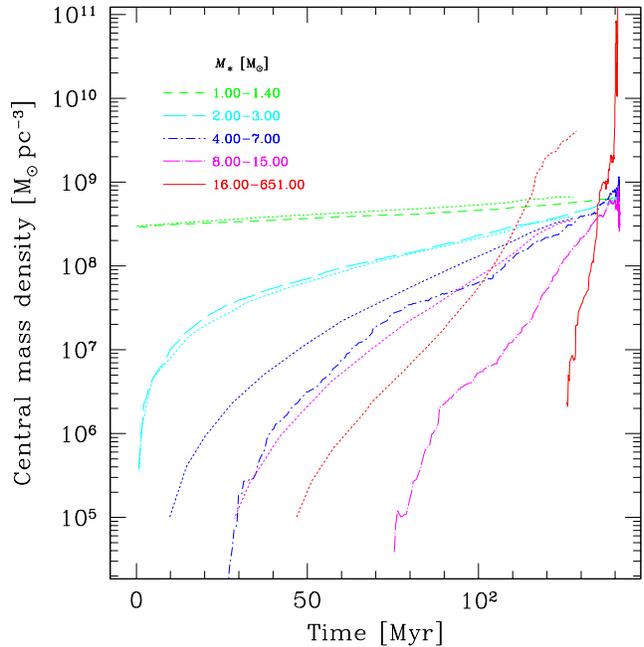}%
        }
\caption{
Mass density of stars of various masses in the central region of a
cluster during its evolution with stellar collisions. Same simulation
as in Figs~\ref{fig:QS90E4A_LargRad} and
\ref{fig:QS90E4A_Nstars} (\Sim{QS90-E4A}). For the $1-1.4\,\Msun$, 
$2-3\,\Msun$, $4-7\,\Msun$, $8-15\,\Msun$ and $\le 16\,\Msun$ mass
bin, we plot the average density within a sphere containing,
respectively, a fraction $0.001$, $0.01$, $0.01$, $0.1$ and $0.1$ of
the total mass in the corresponding range of stellar masses. As in
Fig.~\ref{fig:QS90E4A_Nstars}, thin dotted lines indicate results of
QS90 (their Fig.~2b). Note that the curve from QS90 closest to our line for
the $4-7\,\Msun$ bin is for their $8\,\Msun$ mass component.}
\label{fig:QS90E4A_CtrDens}
\end{figure}

\begin{figure}
  \resizebox{\hsize}{!}{
          \includegraphics[bb=32 148 563 690,clip]{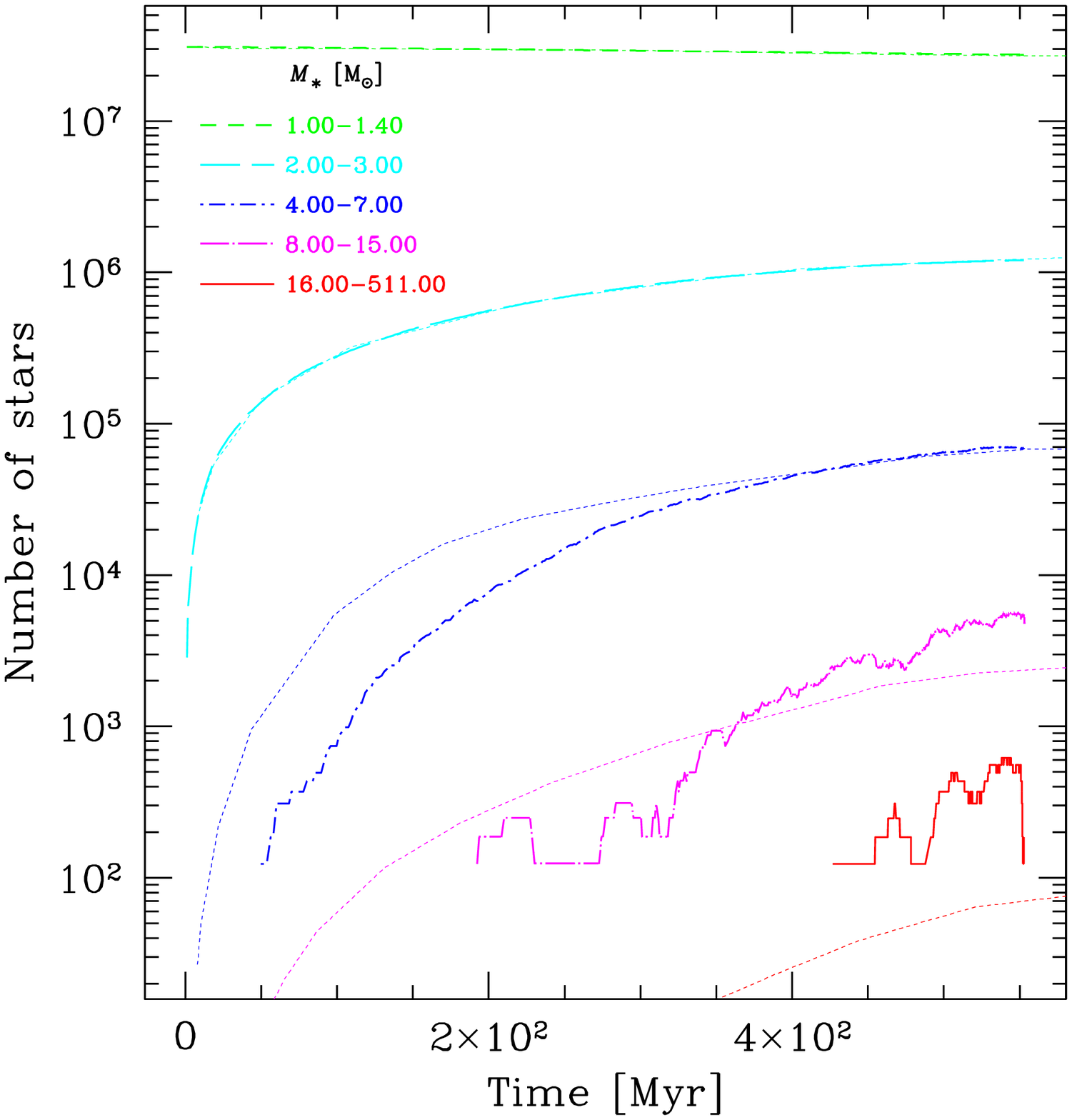}%
        }
\caption{
Number of stars in various mass bins during cluster evolution with
stellar collisions. Our simulation for model E4B (\Sim{QS90-E4B}) 
is compared to data from QS90 (their Fig.~4a). See
caption of Fig.~\ref{fig:QS90E4A_Nstars} for explanations.}
\label{fig:QS90E4B_Nstars}
\end{figure}

In \citet{FB02b}, we have already checked that {\MESSY} produces
the correct rate of collisions, both as a function of the distance to
the centre of the cluster and of the masses of stars, for multi-mass
clusters. We have also reproduced the overall results of collisional
models of MBH-hosting galactic nuclei considered by \citet{DS83} and
\citet{MCD91}. In such environments, however, collisions occur at very 
high relative velocities and do not lead to runaway growth \citep{FGR04c}.

To our knowledge,
\citet[][hereafter QS90]{QS90}, using a FP code, were the first to
study in a systematic way the process of collisional runaway
accounting self-consistently for cluster dynamics and secular
evolution and how collisions themselves affect it.

We have run simulations for all 6 ``E'' models of QS90, assuming, in
most cases, sticky-sphere collisions and complete rejuvenation for
each merger as these authors did. Initial conditions are high-density
Plummer clusters with all stars of solar type, with various total
masses and sizes. Table~\ref{table:QS90} lists the characteristics of
these models. Stellar evolution is included. QS90 implemented it in a
simplified way, similar to ours. However, while we can assign an
individual age to each MC particle (which represents 4.2 to 36 stars,
depending on the model), in the FP scheme of QS90 only a small number
of discrete mass component are present (for stars of mass $2^i\,\Msun$
with $i=1,2,\ldots 8$), each of which represents some homogeneous
population. Stellar evolution can only be treated in statistical way,
with each component having an average age. Similarly, collisions can
only be accounted for statistically by integrating merger rates within
and between mass components and distributing the number of merger
products over components in a way that conserve total mass but not
necessarily individual stellar masses because not all possible merger
masses are represented. Unlike QS90, we have allowed for collisions
between MS stars and compact remnants by assuming complete disruption
of the MS star but no change to the mass or velocity of the compact
object. Certainly an oversimplification, this prescription obviously
corresponds to the most unfavourable possible one, as far as
collisional runaway is concerned. To have the same ratios between the
time scale for relaxation and that for other processes (collisions,
stellar evolution) as QS90, we adopt their $\GCoulomb=0.4$ value for
this set of simulations.

In contrast to QS90 we find core collapse and collisional runaways in
all cases but E2B whose core contraction is stopped by the mass
loss due to stellar evolution of collision products.

Agreement for model E4A is very satisfying: in our model
\Sim{QS90-E4A}, we find core collapse and runaway to happen at
${\tcc}\simeq 140\,{\rm Myr} \simeq 1.44\,\trh(0)$, to compare with
QS90's value of $\sim 130$\,Myr. Without collisions, a single-mass
Plummer model would experience core collapse at $\tcc\simeq
17-18\,\trh(0)$ so mergers clearly play an important role in the
cluster evolution from the beginning, by dissipating energy and
creating a mass spectrum, hence allowing mass segregation. For this
reason, one cannot predict the occurrence of runaway in those clusters
just from the relaxational value of the core-collapse time,
$\left.{\tcc}\right|_{\rm rlx}$. Also, even though all stars are
initially of one solar mass, it is not sufficient to have
$\left.{\tcc}\right|_{\rm rlx}< 10\,$Gyr for collisional runaway to
kick in. Paradoxically, if merger products are formed before deep
collapse they may be able to stop collapse as they evolve off the MS
much earlier, as happens in our case E2B (\Sim{QS90-E2B}).

Fig.~\ref{fig:QS90E4A_LargRad} shows the evolution of the
Lagrange radii our simulation of case E4A. We have also redone the
simulation with the more realistic prescriptions for collision
outcomes based on SPH simulations (\Sim{QS90-E4Asph}). As can be seen
in Fig.~\ref{fig:QS90E4A_LargRad}, when collisional mass-loss and,
more importantly, fly-by, non-merging collisions are accounted for,
the cluster evolution is significantly slower. This is because, in
clusters with such a high velocity dispersion (${\sigmaOneD}_{,\rm c}(0)\simeq
231\,\kms$), only a minority of collisions result in mergers,
amounting to a reduction in the effective collision cross sections. In
Fig.~\ref{fig:QS90E4A_Nstars}, we have plotted the evolution of the
number of stars of various masses during the evolution of
\Sim{QS90-E4A}. The evolution of the number of stars of mass 2 
to $3\,\Msun$ in our run is nearly identical to that of the $2\,\Msun$
component of QS90 but we obtain significantly fewer stars more massive
than $3\,\Msun$. This demonstrates that the rate of collision
between $1\,\Msun$ (and hence, of formation of first-generation
mergers) is the same in both simulations. Because QS90's FP method
imposes a completely different (and much less physical) way of dealing
with mergers of mergers, it is not surprising no close agreement is
reached on that matter. The $M$--$R$ relation cannot be blamed,
however, as they assume $R_\ast/\Rsun = (M_\ast/\Msun)^{0.55}$, which
is close to our relation. The drop in the number of stars more massive
than $16\,\Msun$ at the end of our simulation corresponds to the
run-away growth of a VMS by merging between massive stars. When we
stopped our run, the VMS had reached $651\,\Msun$. We refer to
Paper~II for a discussion of physical processes that may terminate the
VMS growth, most of which are lacking from {\MESSY}. How mergers
create a population of massive stars which come to dominate the
central region of the cluster is further indicated by
Fig.~\ref{fig:QS90E4A_CtrDens}. Here, we follow the contribution to
the central density of stars in various mass ranges. Unlike QS90, we
cannot determine the density at $R=0$ but have to sum the mass in some
small spherical volume to estimate it. Our values are therefore lower
limits. Still, the agreement for the lowest two mass bins
($1-1.4\,\Msun$ and $2-3\,\Msun$) is very satisfying, but again, we
get considerably fewer higher-mass stars.

Although 3 times less massive, model E4B has the same velocity
dispersion as E4A and, hence a very similar initial value of
$\trlx/\tcoll$ (up to small differences in Coulomb logarithms). Therefore,
one could have expected collisions to play the same role and to get
the same value of $\tcc/\trh(0)$. However, stellar evolution introduce
still another time scale in the problem and because $\trh(0)$ is
longer (in years) for model E4A, more collision products have time to
evolve off the MS, which delays core collapse. At $t=0.13\,T_{\rm
FP}$, the fractional number of white dwarfs in our E4A run is $\sim
10^{-5}$ while it reaches $\sim 10^{-3}$ in run E4B. In QS90's run for
this case, stellar evolution was able to stop core collapse around
$t\approx 700\,$Myr, before any star more massive than $32\,\Msun$
formed. In contrast, our models produced a runaway merger after
$520$ to $570\,$Myr already. The evolution of number of stars of different
masses is plotted in Fig.~\ref{fig:QS90E4B_Nstars} and compared with
QS90's results (for the contraction phase). Again, the
agreement with QS90 is excellent for the $2-3\,\Msun$ mass bin but
poorer for others. This time, our run produces more massive stars
(after $t\approx 300\,$Myr), which eventually leads to
runaway. Clearly, because the evolution is driven by a positive
feedback loop between mass segregation and collisions and because only
stellar evolution of merger products themselves can revert it,
differences in the treatment of collisions may have very important
effects.

For their other ``E'' models, QS90 only indicate the end state in a
very succinct way, giving no details about the evolution. For model
E2A, we find runaway at $t\simeq 400\,$Myr, to be compared with QS90's
value of $t\simeq 490\,$Myr, a satisfying agreement, given how
different the approaches are. Model E2B is prevented from entering the
run-away phase by stellar evolution in both QS90's and our
simulation. In our simulation (\Sim{QS90-E2B}), a few particles grow
to $8\,\Msun$ and one to $11\,\Msun$ during the phase of central
concentration. QS90 report that the most massive stars formed have
$8\,\Msun$ in this case. In QS90, the core collapse of model E1A is
reversed by the formation of, and energy input from 3-body binaries, a
process we do no take into consideration (see discussion in
Appendix~\ref{sec:binaries}), but the fact that they form stars as
massive as $64\,\Msun$ indicates that they nearly reach the conditions
for runaway. For this cluster we obtain runaway at $t\simeq
800\,$Myr. Finally, E1B is another case for which QS90 find the core
collapse is interrupted by stellar evolution but our simulation
produces runaway at $t\simeq 0.37$\,Gyr. In one of our simulations for
this model, the runaway star is destroyed in a collision with a
compact object, just after it has reached a mass of
$157\,\Msun$. Another runaway sequence starts shortly afterwards.

Since QS90, runaway collisions in clusters have been studied by use of
the direct $N$-body method
\citep{PZMMMH99,PZMcM02,PZBHMM04}. Unfortunately, these simulations 
were either limited to $\Npart \lesssim 10^5$, in which case binaries
play a crucial role or, for the few cases reaching $\Npart \simeq
6\times 10^5$ and experiencing runaway, done for clusters with high
initial concentration (King parameter $W_0\ge 9$) and, hence probably
dominated by small-number effects in the innermost regions. In
Paper~II, we report briefly on the simulations we have done with initial
conditions similar to those of some models of \citet{PZBHMM04}.


\section{SUMMARY}

The results presented here are mostly tests to ensure that the Monte
Carlo stellar dynamics code we use, {\MESSY} \citep{FB01a,FB02b},
correctly treats the key processes at play in the runaway route, which
are:
\begin{enumerate}
 \item Mass segregation-induced core collapse, driven by 2-body
 relaxation in cluster with a broad, realistic mass function (Salpeter
 of Kroupa, with $M_{\ast,\rm min}=0.1-0.2\,\Msun$ and $M_{\ast,\rm
 max}\simeq 100\,\Msun$) \label{item:relax}
\item Effects of collisions in the evolution
 of the cluster and occurrence of collisional runaway. \label{item:coll}
\end{enumerate}

This paper is a companion to Paper~II \citep{FGR05} in which we
present the results of our large set of simulations to determine the
conditions for, and characteristics of collisional runaway in young
stellar clusters. In the scenario we investigate, the collisional
phase is brought up by the concentration, through relaxation, of
massive stars in the centre of a cluster. A star much more massive
than $100\,\Msun$, formed in a quick sequence of collisions may not
only be a progenitor for an IMBH but is in itself an exotic object of
considerable interest.

To address point~\ref{item:relax}, we performed simulations of
clusters with {\MESSY}, {\NBFOUR} and the gaseous-model code
{\SPEDI}. The only physical process included in the MC and
gaseous-model runs was 2-body relaxation, treated in the standard
Fokker-Planck approximation, i.e., as sum of uncorrelated small-angle
two-body scatterings whose rate is determined by local conditions at
each point in the cluster. The $N$-body code treats Newtonian
gravitation in an essential exact way, without assumptions about the
nature of relaxation. {\MESSY} produces results in close agreement
with their $N$-body counterparts for $\mu=M_{\ast,\rm max}/\langle
M_{\ast}
\rangle \le 20$ at least. Realistic IMF correspond to $\mu \gtrsim
100$, however. There are clear discrepancies between the MC and
$N$-body simulations in this regime. Most noticeably $N$-body results
show an initially stronger but more progressive concentration of
massive stars in the central regions. Nevertheless, the most important
characteristics of the core collapse as a path to the collisional
runaway stage, namely the time for it to happen, $\tcc$, and the
magnitude of central mass segregation during deep collapse are similar
in both type of simulations. Furthermore, {\SPEDI} yields results 
very close to those of {\MESSY}, except for a value of $\tcc$ 13\,\%
shorter.

Turning now to point~\ref{item:coll}, we note that only very few
numerical simulations of the evolution of clusters subject to
relaxation and collisions, up to and including the runaway stage have
been published. Putting aside the recent $N$-body runaway simulations
\citep{PZMMMH99,PZMcM02,PZBHMM04}, not suitable for clear-cut comparison 
with {\MESSY} results as small-$N$ effects may play a strong role in
them, we are left with the older FP simulations of
\citet[][QS90]{QS90}. These models lack realism because they start with a 
single-mass population of $1\,\Msun$ stars. Collisions are assumed to
result in perfect mergers, with no mass loss. In most cases considered,
the cluster is dense enough to promote stellar mergers at early
times. The more massive stars thus formed accelerate collapse by mass
segregation, hence further increasing the merger rate. These same
merger products can also terminate the process of central density
build-up through their mass loss when they evolve off the MS, well
before the original $1\,\Msun$ would have done so. 

We have simulated the 6 models of QS90 for which these authors have
assumed that the gas lost through stellar evolution escapes the
cluster (rather than staying in it and forming new stars). Although
collisional runaway happens more often in our simulations than in QS90
runs, we obtain good general agreement with the relatively scarce data
published by QS90. For the two situations in which they report core
collapse uninterrupted by stellar evolution or 3-body binaries, we
obtain core collapse times 8\,\% longer and 18\,\% shorter than
theirs, which we find satisfying considering the widely different
numerical methods and, especially, treatments of collisions, of
critical importance here. The production rate of $2-3\,\Msun$ objects
in our runs is nearly exactly the same as for QS90's $2\,\Msun$ mass
bin but noticeable differences exist for higher-mass objects, which,
resulting from longer merger sequences, are more affected by
differences in treatments of collisions. For a large part, the
discrepancies between ours and QS90's certainly originates in
this. Our treatment of collisions being much more direct and accurate
than the one allowed by the FP code of QS90, the differences found in
these comparisons do not cast doubt on the ability of {\MESSY} to deal
with collisional cluster evolution.  

We have also re-simulated one of the QS90 models with the highest
velocity dispersion using our more realistic treatment of collisions,
based on SPH simulations
\citep{FB05} and found a significantly longer core-collapse time due 
to the collisions being less effective at producing higher-mass
stars. This stresses the importance of using collision prescriptions
which account for fly-bys and mass loss, in the high-velocity
regime. This question is investigated in more depth in Paper~II.

\appendix


\section{Neglect of binaries}
\label{sec:binaries}

\begin{figure}
  \centerline{\resizebox{0.9\hsize}{!}{ \includegraphics[bb=76 158 459
  689]{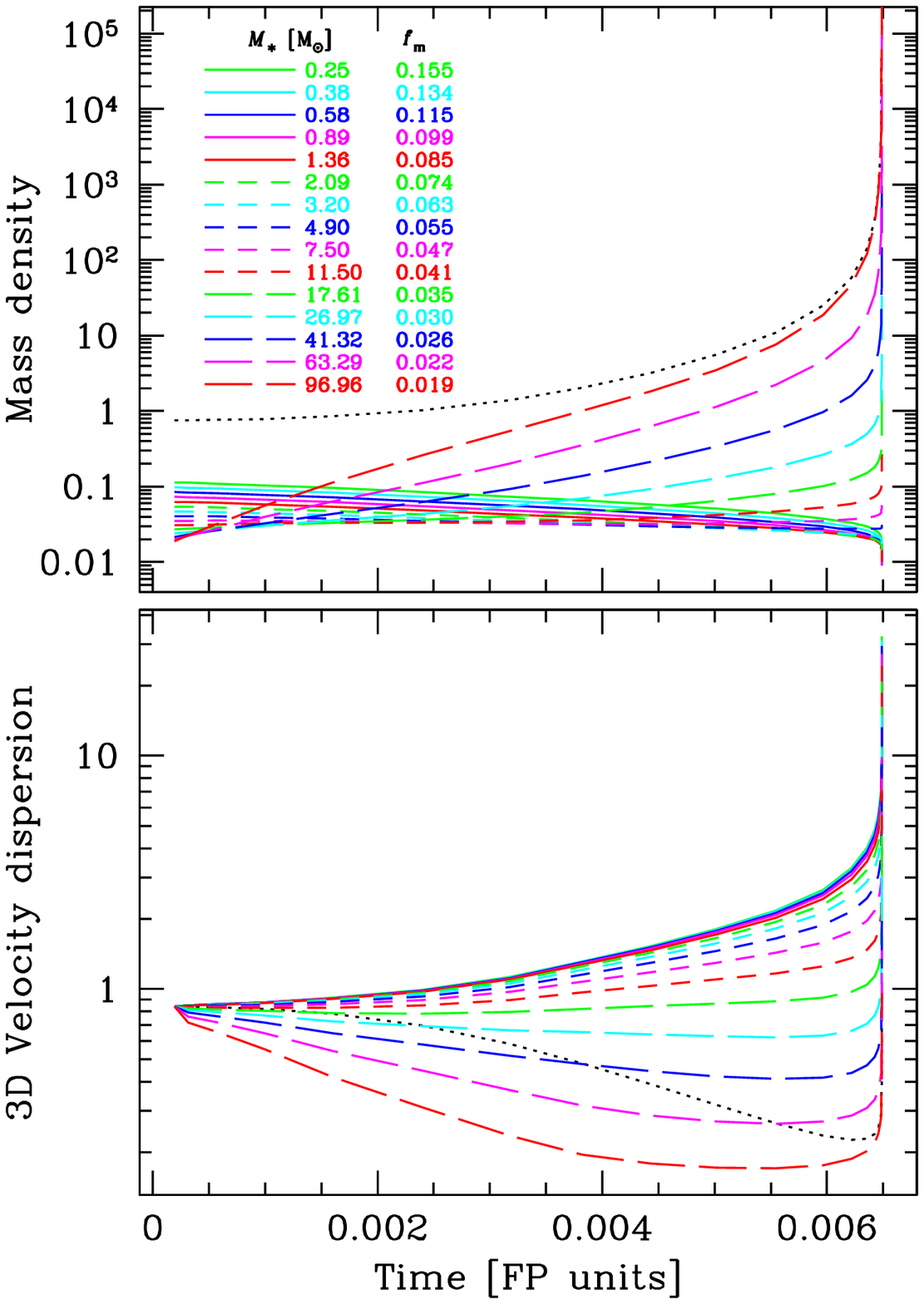} }}

  \caption{ Core collapse of a $W_0=3$ King cluster with Salpeter mass
  function, computed with {\SPEDI}. The mass spectrum is
  discretized into 15 components with the indicated individual stellar
  masses ($m$) and mass fraction ($f_{\rm m}$). On the top panel, we
  represent the contribution of each mass component to the central
  mass density. The dotted line is the total density. The bottom panel
  shows the velocity dispersion of each component. The
  dotted line is the mass-averaged dispersion.}
  \label{fig:spedi_ctr}
\end{figure}

\begin{figure}
  \centerline{\resizebox{0.9\hsize}{!}{%
    \includegraphics[bb=76 158 459 689,clip]{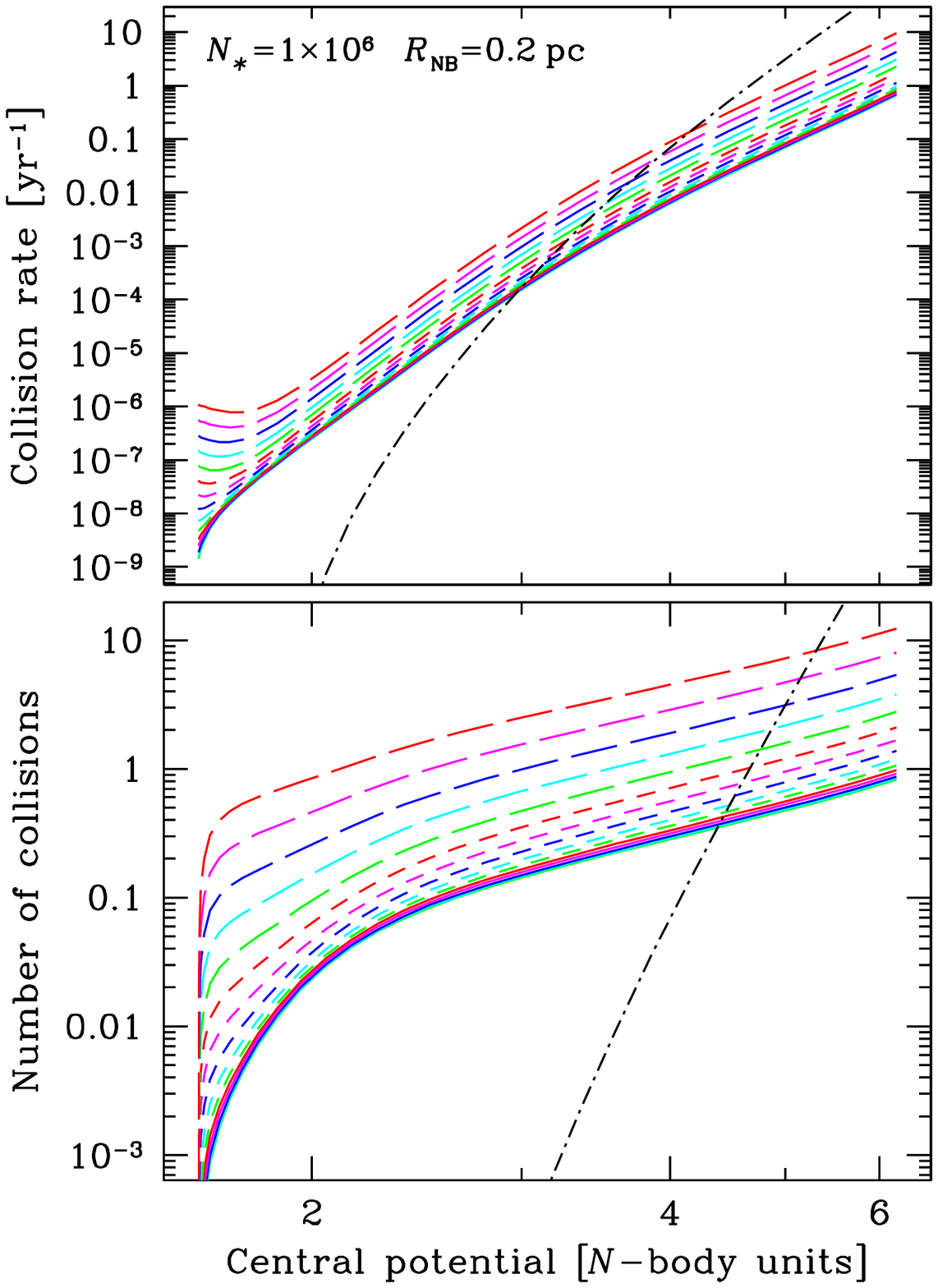}%
    }}
  \caption{%
  Same gaseous model simulation as on
  Fig.~\ref{fig:spedi_ctr}. Here we plot, for each mass component,
  the central collision rate per star (top panel) and integrated
  number of collision or collision probability (bottom panel) for a
  star near the centre. To avoid double counting, the rate for
  component $i$ includes collisions with all components $j\le i$. The
  steep (black) dash-dotted line is an estimate of the formation rate
  of 3-body binaries (per star) obtained by using average stellar
  density and mass in equation~(\ref{eq:3bb_rate}). Note that this
  simulation do not include the effects of collisions or binary
  formation but only 2-body relaxation.  Collision and binary
  formation rates have been estimated {\em a posteriori} assuming the
  cluster is made of $10^6$ stars and its size
  is $\Rnb=0.2\,$pc.}
  \label{fig:spedi_ctr_coll}
\end{figure}

In principle, ``hard'' binary stars, either primordial or formed
during core collapse through 3-body processes, may play an important
role in the cluster evolution. During gravitational encounters with
single stars, a hard binary is likely to shrink, thus allowing the
single star (possibly a former member of the binary if an exchange has
occurred) to emerge with increased kinetic energy. Through this
dynamical heating, hard binaries may suspend or even reverse core
collapse \citep[][and references therein]{HH03}. Obviously, this might
prevent the core from ever entering the high-density collisional phase
needed in the runaway scenario. However, when the finite size of stars is
taken into account in the numerical study of single-binary and
binary-binary interactions, it appears that the collision of at least
two of the stars is a likely event (see \citealt{FregeauEtAl04} for
the most recent cross-section computations and references about this
question). Although using point-mass dynamics, \citet{GS03} studied
the statistics of binary-single and binary-binary encounters occurring
in their simulations of clusters containing primordial binaries and
found that of order 50\,\% of these interactions should indeed lead to
stellar mergers, for typical globular cluster parameters. It is
therefore possible that binaries actually {\em foster} collisions
rather than preventing them. This is exactly what was found in $N$-body
simulations of relatively small clusters (most of them with ${\Nstar}\le
131\,072$, one run with ${\Nstar} \simeq 585\,000$) by Portegies Zwart and collaborators
\citep{PZMMcMH99,PZMcM02,PZBHMM04}. Furthermore, dissipative processes, such as
collisions or tidal interactions, occurring during binary interactions
may significantly reduce the heating effect of hard binaries, compared
with the point-mass approximation \citep{McMillan86,GH91,CH96}. To our
knowledge, however, the impact of this on core collapse has not
yet been studied explicitly.

Even if binaries probably do not prevent core collapse and collisions,
it does not follow that they will not impede or modify the runaway
process. Indeed, the cross section for collision of a single star with
a binary is of order $\pi G a M ({\Vrel})^{-2} \propto M$ where $a$ is
the binary semi-major axis and $M$ the mass of the three stars,
while it is approximately $\pi G R M ({\Vrel})^{-2} \propto
M^{\alpha}$ with $\alpha\simeq 1.5$ for the collision with a single
more massive star of mass $M$ ($>50\,\Msun$) and radius $R$. According
to mathematical modelling through the coagulation equation, runaway is
expected only if the cross section scales like $M^\zeta$ with
$\zeta>1$ \citep{LeeMH93,LeeMH00,MG02}. This condition is not obeyed
in the case of binaries competing with each other for interaction and
merger with single stars. However, the coagulation equation is clearly at
best a crude idealisation for the complex stellar dynamical situation
of interest here.


To simplify the problem, we have assumed in this work that no
primordial binaries were present. We now examine if we are then
justified to neglect binary process altogether consistently, in spite
of the possibility of forming binaries dynamically. We do not consider
binary formed by tidal interactions because their formation requires
close passage at a distance of a few stellar radii at most. Their
formation rate cannot then be significantly higher than collision rate
and their post-capture evolution, though still a controversial issue,
is likely to lead to merger anyway. We instead consider the question
of binaries formed through (point-mass) 3-body interactions.

The creation rate of such binaries is (\citealt{BT87}, Eq.~[8-7]; see
also appendix of \citealt{IBFR05})
\begin{equation}
\dot{n}_{\rm 3b} \simeq C_{\rm 3b} n^3
\frac{G^5 M_\ast^5}{\sigmaOneD^9}, 
\label{eq:3bb_rate}
\end{equation}
where $n$ is the stellar number density and $C_{\rm 3b}\simeq 0.75$ \citep{GH93,HH03}. 

We compare this with the collision rate,
\begin{equation}
\begin{split}
\dot{n}_{\rm coll} = n t_{\rm coll}^{-1} & \simeq  
4 \pi n^2 R_\ast^2 \left(1+\frac{2 G M_\ast}{R_\ast
    ({\Vrel})^2}\right) {\Vrel} \\
 & \approx 8\pi \frac{n^2 G M_\ast R_\ast}{\sigmaOneD}.
\end{split}
\end{equation}
One gets
\begin{equation}
\frac{\dot{n}_{\rm coll}}{\dot{n}_{\rm 3b}} \approx 
\frac{8\pi}{C_{\rm 3b}} \frac{R_\ast \sigmaOneD^8}{G^4 M_\ast^4 n} \approx
\frac{500}{n R_\ast^3} \left(\frac{\sigmaOneD}{\Vstar}\right)^8 
\end{equation}
with $V_\odot=(2G\Msun/\Rsun)^{1/2}=618\,\kms$. For typical
$100\,\Msun$ stars (dominating the central regions), $\Rstar\simeq
14\,\Rsun$, $\Vstar\simeq 1670\,\kms$ and
\begin{equation}
\left.\frac{\dot{n}_{\rm coll}}{\dot{n}_{\rm 3b}}\right|_{100\,\Msun} \approx 
7  \left(\frac{n}{10^6\,{\rm pc}^{-3}}\right)^{-1} \left(\frac{\sigmaOneD}{20\,\kms}\right)^{8}.
\end{equation}
Thus, one cannot clearly exclude that the formation of 3-body binaries
will not compete with direct collision, at least for systems with a
relatively low velocity dispersion. 

Estimating {\it a posteriori} the central values of $\dot{n}_{\rm 3b}$
from MC runs is very difficult because of the steep dependences on $n$
and $\sigmaOneD$, which make the estimate extremely noisy. Thus, to see
how how $\dot{n}_{\rm coll}$ and $\dot{n}_{\rm 3b}$ evolve during core
collapse, we resort to the {\SPEDI} gas-model code presented in
Section~\ref{subsec:spedi}. In Figure.~\ref{fig:spedi_ctr}, we follow
the evolution of central densities and velocity dispersions in a
15-component {\SPEDI} core-collapse simulation. Assuming
${\Nstar}=10^6$ and ${\Rnb}=0.2\,{\rm pc}$, we can now compute what
the central collision and 3-body formation rates would have been
during the core collapse.  Because the evolution speeds up near the
moment of core collapse, we use the central potential instead of time
as independent variable in Fig.~\ref{fig:spedi_ctr_coll} where we plot
the instantaneous and time-integrated rates for all mass
components. For this size and star number, the cluster should become
collisional before the first binary forms. How this depends on the
cluster parameters is expressed by either of the scalings
\begin{equation}
  \frac{\dot{n}_{\rm coll}}{\dot{n}_{\rm 3b}} \propto {\Nstar}^3 R_{\rm cl}^{-1}
  \propto \left(\ln\Lambda\,T_{\rm rh}\right)^{-2/3} {\Nstar}^{10/3}
\end{equation}
where $R_{\rm cl}$ is some characteristic cluster size (e.g., the
half-mass radius). The strong dependence on ${\Nstar}$ suggests that
3-body binaries, which appear to dominate the collision process in
$N$-body simulations may be of little importance in larger systems
with ${\Nstar} \gtrsim 10^6$, such as very massive young clusters or
proto-galactic nuclei. This analysis is in agreement with the results
of \citet{PZBHMM04} who find the dynamically-formed binaries to play a
lesser role in comparison with earlier, lower-N simulations
\citep{PZMcM02}.

\section*{Acknowledgements}

We thank Melvyn Davies, John Fregeau, M.~Atakan G\"urkan, Jamie Lombardi,
Simon Portegies Zwart, Rainer Spurzem and Hans Zinnecker for useful
discussions. We are grateful to Rainer Spurzem for making the code for
gaseous models ({\SPEDI}) available and for his assistance in using
it. MF is indebted to Pau Amaro Seoane for his extraordinary support
during his stay in Heidelberg. The work of MF was funded in part by
the Sonder\-forschungs\-bereich (SFB) 439 `Galaxies in the Young
Universe' (subproject A5) of the German Science Foundation (DFG) at
the University of Heidelberg. This work was also supported by NASA ATP
Grants NAG5-13236 and NNG04G176G, and NSF Grant AST-0206276 to
Northwestern University.

\label{lastpage}

\end{document}